\documentclass[]{article}
\usepackage[top=4cm, bottom=4cm, left=1.25in, right=1.25in]{geometry}
 \usepackage{varioref}
 \usepackage{wrapfig}
 \usepackage{threeparttable}
 \usepackage{amsmath}
 \usepackage{amssymb}
 \usepackage{bm}
 \usepackage{dcolumn}
  \newcolumntype{d}{D{.}{.}{-1}}
 \usepackage{nomencl}
 \makenomenclature
 \usepackage{subfigure}
 \usepackage{subfigmat}
 \usepackage{fancyvrb}
  \fvset{fontsize=\footnotesize,xleftmargin=2em}
 \usepackage{lettrine}
 \usepackage{tabularx}
 \usepackage{color}

 \usepackage{./myplots}
 \usepackage{commath}
 \usepackage{tikz}
\usepackage{placeins} 

\usepackage{soul} 
\usepackage[normalem]{ulem} 

\usepackage{natbib}

\usepackage{xcolor}
\newcommand{\reva}[1]{\textcolor{black}{#1}}

\author{Alejandro Gonzalo$^{1}$, Manuel García-Villalba$^{2}$ and Oscar Flores$^{3}$ \\[2ex]
$^1$ Department of Mechanical Engineering,\\ University of Washington, Seattle, WA, United States\\[1ex] 
$^{2}$ Institute of Fluid Mechanics and Heat Transfer\\ TU Wien, Vienna,  Austria \\[1ex]
$^{3}$ Aerospace Engineering Department\\ Universidad Carlos III de Madrid, Spain \\[1ex] 
\small Corresponding author: manuel.garcia-villalba@tuwien.ac.at
 }

\title{A quantitative analysis of leading edge vortices \\ in flapping wing aerodynamics}

\newlength{\twhalf}
\newlength{\twthird}
\newlength{\twfourth}
\definecolor{c01}{rgb}{0.000,0.447,0.741}
\definecolor{c02}{rgb}{0.850,0.325,0.098}
\definecolor{c03}{rgb}{0.929,0.694,0.125}
\definecolor{c04}{rgb}{0.494,0.184,0.556}
\definecolor{c05}{rgb}{0.446,0.674,0.188}
\definecolor{c06}{rgb}{0.301,0.745,0.933}
\definecolor{c07}{rgb}{0.635,0.078,0.184}
\definecolor{c11}{rgb}{0.000,0.000,1.000}
\definecolor{c12}{rgb}{0.000,0.000,0.000}
\definecolor{c13}{rgb}{1.000,0.000,0.000}
\definecolor{c14}{rgb}{0.000,0.500,0.000}

\newcommand{\solid}[1]{\lineSymbol[solid]{#1}{0.018in}{none}{0in}}
\newcommand{\dash}[1]{\lineSymbol[dashed]{#1}{0.018in}{none}{0in}}

\begin{document}

\maketitle

%
\begin{abstract}
The leading edge vortex (LEV) is one of the most important
lift augmentation mechanisms in flapping wing aerodynamics.
We propose a methodology that aims to provide a quantitative description of the LEV.
The first step of the method consists of 
the identification of the vortical structures surrounding the wing using the $Q$ criterion.
The impact of the employed threshold is shown to be minor, not influencing the observed trends.
In the second step we identify the core of the LEV using a thinning algorithm,
discriminating the LEV using the orientation of the locally averaged vorticity vector.
Finally, we compute relevant flow quantities along the LEV core,
by averaging in planes perpendicular to the local vorticity at the LEV core points. 
We have applied this methodology to flow data corresponding to 
a pair of wings performing a flapping motion in forward flight at moderate Reynolds number. 
We have performed a geometrical characterization of the 
LEV and we have computed several flow quantities along the LEV core. 
For the particular configuration under study, we have shown that the LEV,
during the first half of the downstroke develops and grows increasing its circulation smoothly. 
Approximately at mid-downstroke the leading edge vortex starts splitting and its downstream part
is advected towards the wake while keeping its circulation rather constant.
Finally, we have briefly explored  the link between the sectional lift on the  wing and the local circulation obtained with the present methodology.

\end{abstract}


%
\section{Introduction}

In recent years there is a growing interest in the development of bio-inspired micro air vehicles (MAVs) 
that mimic the flight of insects and small birds \citep{haider2021recent}. 
Such development efforts require a deep understanding of the unsteady aerodynamic mechanisms behind their flapping flight,
in order to guide the design of MAVs able to exploit these mechanisms for enhanced maneuverability.
These unsteady mechanisms were identified rather early \citep{ellington1984d}
and are well known: leading edge vortex (LEV), clap-and-fling, wake capture, etc \citep{shyy2013,liu2024vortices}.
Among them, probably the most important one is the LEV \citep{eldredge2019leading}
In spite of this, it has proven difficult to use this knowledge for the systematic design of MAVs. 
Probably the reason behind is that, at present, the acquired knowledge is mostly of a qualitative nature.
For example, it is well known that an attached LEV provides 
additional lift \citep{ellington1996,dickinson1999}, however, 
it is difficult to predict exactly how much additional lift 
is obtained as a function of the large number of parameters involved in flapping flight.
Furthermore, for a general flapping motion, the LEV might present a non-trivial 3D structure,
whose precise geometry, location and intensity is very difficult to predict a priori.

In the literature, several approaches are used to identify the LEV.
For instance, \citet{visbal2011b,visbal2011a} used pressure contours to qualitatively identify vortices as regions of low pressure.
\citet{visbal2013} determined some specific
features of the LEV (e.g. the motion of the legs of the arch-vortex) by means of the
phase-averaged surface pressure.
Other authors  employ the vorticity field to identify the vortices \citep{birch2003,taira2009,Jones2011,jardin2014,calderon2014}.
Thus, regions with high vorticity values are considered vortices,
although it is important to note that this magnitude is also high in shear
layers.
There are other local approaches based on the velocity gradient tensor that do
not present this limitation.
Among others, the $Q$ \citep{hunt1988} and the $\lambda_2$ \citep{jeong1995} criteria are extensively used
in the unsteady aerodynamics field to identify vortical structures \reva{\citep{taira2009,kweon2010,visbal2011b,jardin2012,harbig2013,jantzen2014,harbig2014,kolomenskiy2014,zhang2020formation,lee2022leading,son2022leading,wei2023experimental,chen2023b}}.
All these approaches have been used to explain qualitatively the flow features
observed around the wing, such as the LEV, the tip vortices (TiVs) and the trailing
edge vortex (TEV).

In order to characterize the effect of the LEV on the aerodynamic
forces generated by flapping wings, 
it is important to provide quantitative information in addition to qualitative one.
This includes for example the precise determination of the relative position of the LEV with respect to the wing and the quantification of the LEV intensity. 
Several authors have attempted to tackle this issue. 
\citet{birch2004}
performed experiments on revolving wings at constant angular velocity, observing 
a stable LEV on the wing. 
They integrated the vorticity over wing cross-sections to 
estimate the local circulation around the wing. 
\citet{Jones2011}
performed a similar experiment at a somewhat higher Reynolds number than  \citet{birch2004}. 
They identified the vortices using the vortex identification method of \citet{graftieaux2001}.
They studied the geometry and location of the LEV as a function of time
by analyzing several wing cross-sections. 
The growth of the LEV at a given cross-section
was quantified by computing the local circulation as a function of time. 
\citet{jardin2014} performed direct numerical simulations
of flow over a wing undergoing different maneuvers. 
They characterized the LEV by analyzing the midspan plane
and computing the local circulation as a function of the distance 
traveled by the wing. 
\citet{jantzen2014} performed direct numerical simulations and experiments of flat-plate rectangular wings undergoing pitching
maneuvers about the leading edge.
They tracked the LEV along the midspan plane  
employing the same vortex identification method \citep{graftieaux2001} as \citet{Jones2011}.
They also evaluated the vortex strength by integrating the spanwise
vorticity inside the vortex core boundary. 
These results were limited to a 2D plane, but the flow visualizations
showed that the LEV was a complex 3D structure, so that different
results might have been obtained at other cross-sections. 
\reva{Lastly, \citet{chen2022} investigated how Reynolds number affects the dynamics and stability of LEV formation in revolving rectangular plates. 
Building on prior knowledge of LEV dynamics in revolving wings, they employed cylindrical slices along the spanwise direction to locate the LEV region and estimate its vorticity and circulation. 
These curved slices are aligned with the wing's motion and the shape of the shed LEV, 
helping to better characterize its intensity. 
This methodology has also been used to characterize the LEV evolution during the rapid rotation of flapping wings \citep{chen2024}.}

In this article, we aim to contribute to the effort of characterizing the LEV.
This will be done by assessing a methodology for providing a quantitative description 
of the LEV that appears on a flapping wing in forward flight. 
The methodology proposed here 
does not rely on the analysis of cross-sections and
takes into account the 3D nature of the LEV.
The paper is organized as follows. A brief description of the numerical database analyzed in this paper is provided in section \ref{sec:cases}. Then, section \ref{sec:methodology} describes in detail the alorithm developed to identify and quantify the LEV. The results obtained from applying this algorithm to the aforementioned database are presented in section \ref{sec:results}, and conclusions are provided in section \ref{sec:conclusions}.  

\section{Computational setup}
\label{sec:cases}

This section provides a brief overview of the computational setup and numerical method 
that were employed to generate the flow data analyzed in this article.
The configuration consists of a pair of wings performing a flapping motion as they
fly forward with velocity $u_\infty$. 
The wings are rectangular with a chord length $c$ and a span $b$, so that the aspect ratio
is $AR=b/c$. 
Additionally the wings are rounded on both inboard and outboard wing tips.
The wings cross-sections consist of NACA0012 airfoils.
The wings are placed side by side, with a separation between their inboard tips of $0.5c$.
The Reynolds number is equal to $Re = u_\infty c/\nu = 500$,
where $\nu$ is the kinematic viscosity.
Each wing rotates with respect to its inboard wing tip with a sinusoidal law of angular frequency $\omega=u_\infty/c$
and a flapping amplitude such that the maximum vertical displacement of the outboard wing tip is $c$.
Note that for simplicity, the wing is not subject to pitching motion.
In this article, flow data corresponding to wings of $AR=2$ and $AR=4$ is analyzed.
The results of the  simulation of $AR=2$ were discussed by \citet{gonzalo:2018}
with emphasis on the characterization of the aerodynamic forces.
The simulation of $AR=4$ was reported in \citet{gonzalo2018aerodynamic}. 

In the discussion, a non-inertial reference frame fixed 
to the wing will be used to study the flow variables.
In that reference frame, $x$ is the chordwise direction, 
$y$ is the spanwise direction and $z$ is the direction 
perpendicular to the mean surface of the wing.
The corresponding unitary vectors along these directions are $\mathbf{e}_{x}, \mathbf{e}_{y}$ and $\mathbf{e}_{z}$, respectively. 
Then, the inboard wing tip is found at $y = 0$ and the outboard wing
tip at $y/c = AR$.
The leading edge of the wing is found at $x = 0$.

The simulations were performed with the in-house code TUCAN, which
solves the Navier-Stokes equations for an incompressible flow and model the
presence of the wings using the immersed boundary method proposed by
\citep{uhlmann2005}.
A detailed description of TUCAN can be found in previous works 
together with extensive validation using both simple test 
cases and comparison with experimental data \citep{moriche2017a,moriche2017b, moriche2020assessing, moriche2021characterization}. 

For the case with $AR=2$, the computational domain size is 
$[12c\times5.25c\times 8c]$ in the streamwise, spanwise
and vertical directions, respectively.
For the case with $AR=4$, the length and height of the domain are the same, 
while the width needs to be increased to $7.25c$.  
In order to save computational time a symmetry boundary condition is imposed 
at the midplane between the wings, and therefore only one wing is simulated.
At the upstream boundary, a uniform free stream is imposed
while a convective boundary condition is imposed at the downstream boundary.
Free slip boundary conditions are imposed at the top, bottom and lateral boundaries.

A uniform mesh was employed in the simulation, with a resolution of $56$ points per wing chord
length in all the spatial directions.
This resolution was determined from a grid convergence study performed in a 2D
simulation of a NACA0012 in heaving motion with an amplitude equal to $c$ 
and the same $Re$ and $\omega$ of the 3D cases described above.
Thus, the total number of grid points is
$N_x = 672$, $N_z = 448$ and $N_y = 294$ ($N_y = 406$) for the case with $AR=2$
($AR=4$).
The simulations were run during several cycles until a periodic
state was reached. Due to the symmetry of the motion, downstroke and upstroke are equivalent, and in the following only the downstroke is analyzed.

%
\section{Methodology}
\label{sec:methodology}

The method to characterize the  LEV proposed here can be summarized in three main steps.
First, the instantaneous vortical structure containing the LEV is identified. 
Second, the skeleton of this vortical structure is determined, allowing for the identification of the position and orientation of the LEV. 
Finally, quantities of physical relevance are averaged as a function of their position along the core of the LEV. 
In the following, these steps are going to be explained in detail.

\subsection{Identification of the vortical structure containing the LEV}

Since the wing kinematics required to perform most of the relevant flight maneuvers
in unsteady aerodynamics (forward flight, hover, perching,...) include
one or more rotations, the flow surrounding the wing is typically studied in a
non-inertial reference frame fixed to it.
Then, the relative velocity $\mathbf{u^\prime}$ at any point of the fluid ($\mathbf{r}$) is
defined as 
\begin{equation}
 \mathbf{u^\prime} = \mathbf{u} - \mathbf{u}_{O^\prime}  -
 \boldsymbol{\Omega} \times (\mathbf{r}-\mathbf{r}_{O^\prime}),
\label{eq:uprime}
\end{equation}
where $\mathbf{u}$ is the absolute velocity of the fluid at $\mathbf{r}$,
$\mathbf{u}_{O^\prime}$ is the velocity of the origin of the non-inertial 
reference frame fixed to the wing ($O^\prime$), $\boldsymbol{\Omega}$ is the 
instantaneous angular velocity of the non-inertial reference frame and $\mathbf{r}_{O^\prime}$ is the position of $O^\prime$.
Taking the rotor of equation (\ref{eq:uprime}), the relative vorticity $\boldsymbol{\omega}^\prime = \nabla \times \mathbf{u^\prime}$ can be related to the absolute vorticity $\boldsymbol{\omega} = \nabla \times \mathbf{u}$, 
\begin{equation}
 \boldsymbol{\omega}^\prime = \boldsymbol{\omega} - 2 \boldsymbol{\Omega}.
\label{eq:omegaprime}
\end{equation}

\begin{figure}
 \begin{center} 
  \begin{tikzpicture}
   \coordinate(O) at (0.,0.); 
   %
   \node(A) at (O) {\includegraphics[width=\twhalf]{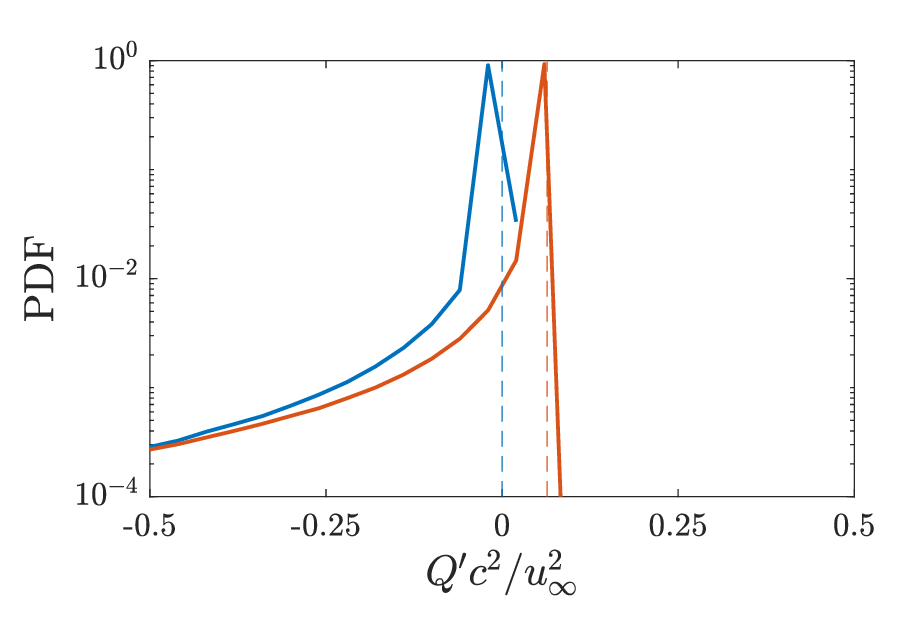}};
  \end{tikzpicture}
  \caption{PDF of  $Q^\prime$  in a 
           volume immediately upstream of the wing (see text for details).
           Two time instants are shown, namely, the
           beginning of the downstroke, $t/T = 0$ (\solid{c01})
           and the mid-downstroke, $t/T = 0.25$ (\solid{c02}).
           Dashed lines (\dash{c01}) and (\dash{c02}) represent the values of
           $\norm{ \mathbf{\Omega} } ^2$ at $t/T = 0$ and
           $t/T = 0.25$, respectively.
  \label{fig:Qth} }
 \end{center}
\end{figure}

In order to define and identify the instantaneous vortical structures, the second invariant 
of the velocity gradient tensor of the relative velocity ($Q^\prime$) is used here \citep{hunt1988}. 
Hence, vortical structures are defined as 3D regions of the flow where
$Q^\prime > Q^\prime_{th}$, as previously used in several works
\citep{taira2009,visbal2011b,visbal2011a,visbal2013,harbig2013,harbig2014,jantzen2014,jardin2017,zhang2020formation,son2022leading}.
Note that in the present case, the choice of $Q^\prime_{th}$ is not trivial. The use of relative velocities imposes a lower bound on $Q^\prime$, which is related to the angular velocity of the wing, $\boldsymbol{\Omega}$. This lower bound is made explicit when $Q^\prime$ is expressed in terms of $Q$, the second invariant of the gradient of the absolute velocity,  
\begin{equation}
 Q^\prime = Q + \norm{ \boldsymbol{\Omega} } ^2 -
            \boldsymbol{\Omega} \cdot \boldsymbol{\omega}.
\label{eq:Qprime}
\end{equation}
The direct consequence is that far upstream from the wing, where the velocity is homogeneous and both $Q$ and $\boldsymbol{\omega}$ are zero, $Q^{\prime} = \norm{ \boldsymbol{\Omega} }^2$.
This is illustrated in Figure \ref{fig:Qth}, where the probability density function of $Q^\prime$ upstream of the wing with $AR = 4$ is compared at two different instants: 
at the beginning of the downstroke ($t/T = 0$) when $\boldsymbol{\Omega}$ is zero, 
and at the mid-downstroke ($t/T = 0.25$) when $\boldsymbol{\Omega}$ is maximum.
It can be observed that at both instants the p.d.f. peaks just before the value of $\|\boldsymbol{\Omega}\|^2$ at that time, indicated by the vertical dashed lines in the figure. The value of the p.d.f. for $Q^\prime > \|\boldsymbol{\Omega}\|^2$ in the region upstream of the wing is essentially zero at both instants. 
Therefore, the maximum value of $\|\boldsymbol{\Omega}\|^2$ during the cycle can be
considered as the minimum value of $Q^{\prime}_{th}$, necessary to avoid the 
identification of spurious vortical structures, generated exclusively by the 
choice of reference frame. 

\def\tmpa{pdf_relQ}
\def\tmpb{2W-AR4-R000}

On the other hand, there is not an {\em a priori} limit on the maximum value of the threshold. Obviously, if the threshold is too high no vortical structures are detected. Hence, a certain range of $Q^\prime_{th}$ needs to be scanned to ensure the robustness of the method, as shown in the next section. 

Finally, it should be noted that, as discussed by 
\citet{chakraborty2005}, the $Q$-criterion is equivalent to other identification
methods (like the discriminant of the velocity gradient tensor, or the swirling
strength) when appropriate thresholds are used. 
The present choice of the $Q$-criterion is based on the fact that this method only requires to calculate low order spatial derivatives of the velocity field 
and products of quadratic order, minimizing the computational resources and the
time required by the algorithm. 
However, it has been checked that the results presented in the next sections are very similar when the $\lambda_2$ criterion is used, with an equivalent threshold in terms of the volume occupied by the identified vortical structures. 

Once a vortex identification method and a threshold has been selected, the next step in the LEV characterization method is to discriminate the vortical structure containing the LEV from other vortical structures in the flow (i.e., mostly the vortical structures shed to the wake in the previous flapping cycle). 
This is accomplished in the present study by computing 
$Q^\prime$ in a volume of fluid surrounding the wing, given by 
$x \in \left[ -0.5c, 2c \right]$, 
$z \in \left[ -1.5c, 1.5c \right]$ and 
$y \in \left[ -1.25c, 3.25c \right]$ for the case with $AR=2$ ($y \in \left[ -1.25c, 5.25c \right]$ for $AR = 4$).
Only the largest coherent object satisfying $Q^\prime > Q^\prime_{th}$ in this volume is kept for the next step. This procedure assumes that the LEV is the largest vortical structure in the region near the wing, which is always true for the present configurations. 
\reva{To illustrate the methodology, the structures at two time instants
for the case with $AR=4$ are analyzed,
namely, $t/T = 0.25$ in figure \ref{fig:FVA} and $t/T = 0.41$ in figure \ref{fig:FVA2}.
In the first instant selected, the vortical structure is quasi-2D and could be studied
with simpler methods such as using cross-sectional cuts. 
In the second instant selected, the vortical structure is more three-dimensional and 
an analysis using cross-sectional cuts is likely to be misleading. 
}
Figures \ref{fig:FVA}$a$ \reva{and \ref{fig:FVA2}$a$} 
show all the vortical structures identified with a threshold $Q^\prime_{th} = 4 u_\infty^2/c^2\gg\| \mathbf{\Omega} \|^2$, while the translucent surface in 
figures \ref{fig:FVA}$b$ \reva{and \ref{fig:FVA2}$b$} correspond to the largest one. 
It is important to note that the latter contains the LEV, but also the TiV around the outboard wing tip and a segment of a TEV. 

\subsection{Identification of the skeleton of the LEV}

Next, the core or skeleton of the vortical structure identified in the previous step is computed. 
This task is done with the thinning algorithm proposed by 
\citet{lee1994} and implemented in MATLAB by \citet{kerschnitzki2013}. 
The algorithm extracts the medial axes centerline of 3-D objects,
preserving their topological and geometrical conditions.
Graphically, the process performed by the algorithm can be described as the
peeling of an onion, being the onion the 3D object satisfying $Q^\prime > Q^\prime_{th}$ and the core of the onion its medial axes centerline.
Voxels (volumetric pixels) at the surface of the vortical structure are discarded, until only the set of points that define the skeleton of the vortical structure are left. 

\def\tmpz{2W-AR4-R000_Qth4000}
\def\tmpa{FVA_isosurface_relQ}
\def\tmpb{FVA_vortex_skeleton_relQ}
\def\tmpc{FVA_dirvecs_relQ_paper_figure}
\def\tmpd{FVA_planes_relQ}
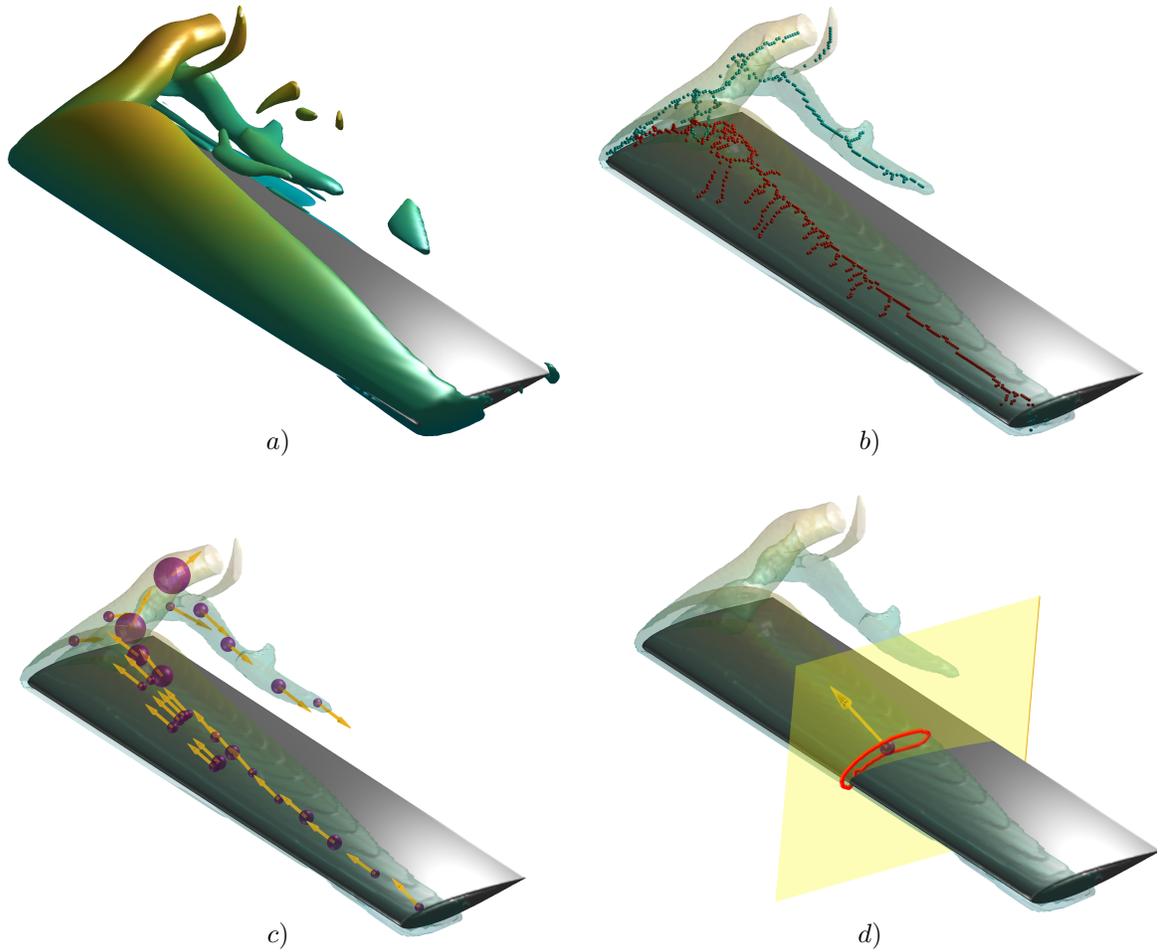
\begin{figure}
 \begin{center} 
  \begin{tikzpicture}
   \coordinate(O) at (0.,0.); 
   %
   \node(A) at (O)              {\includegraphics[width=\twhalf,trim=3.2cm 1.25cm 4cm 1.25cm, clip]{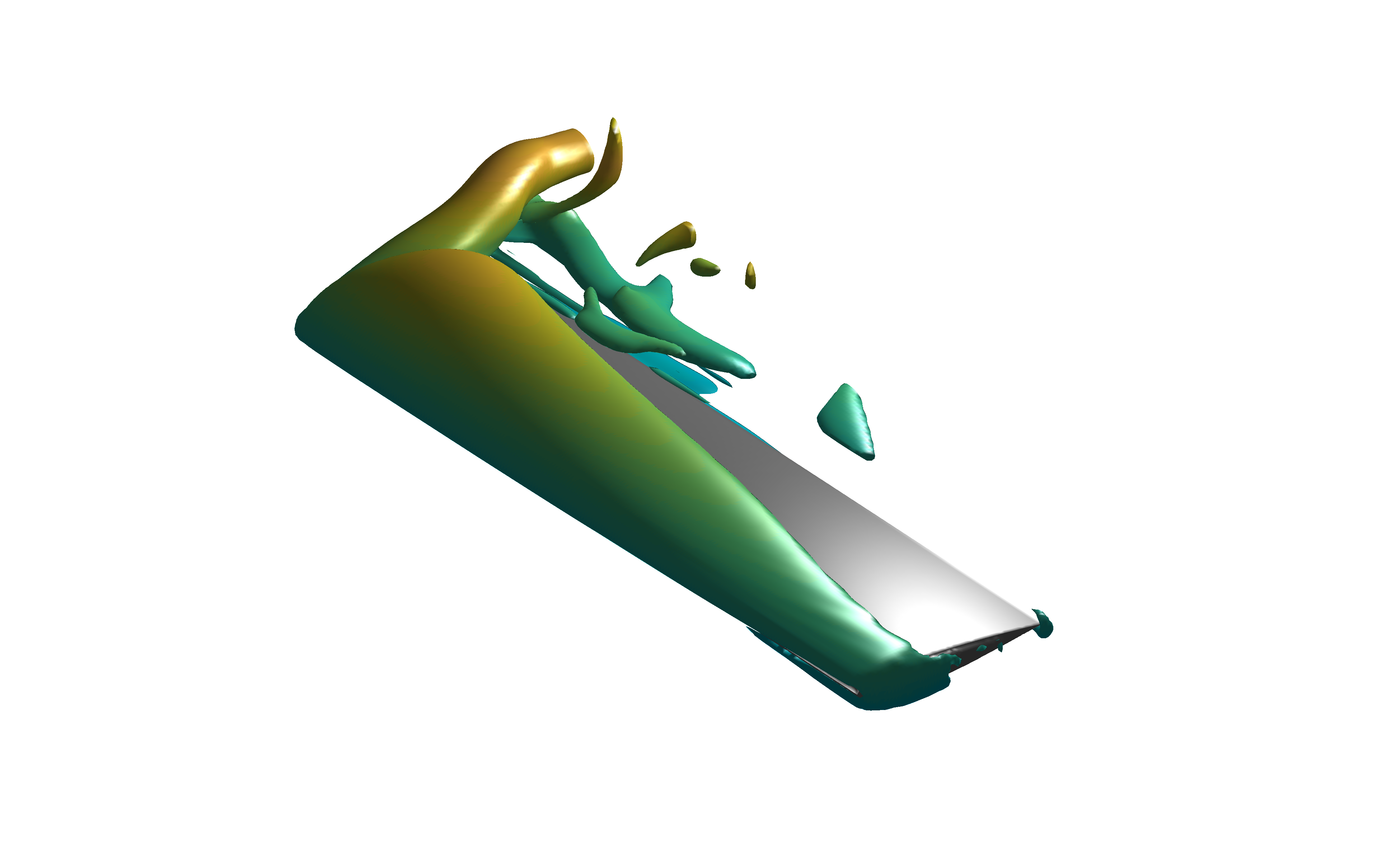}};
   \node(B) at (A.east)[right]  {\includegraphics[width=\twhalf,trim=3.2cm 1.25cm 4cm 1.25cm, clip]{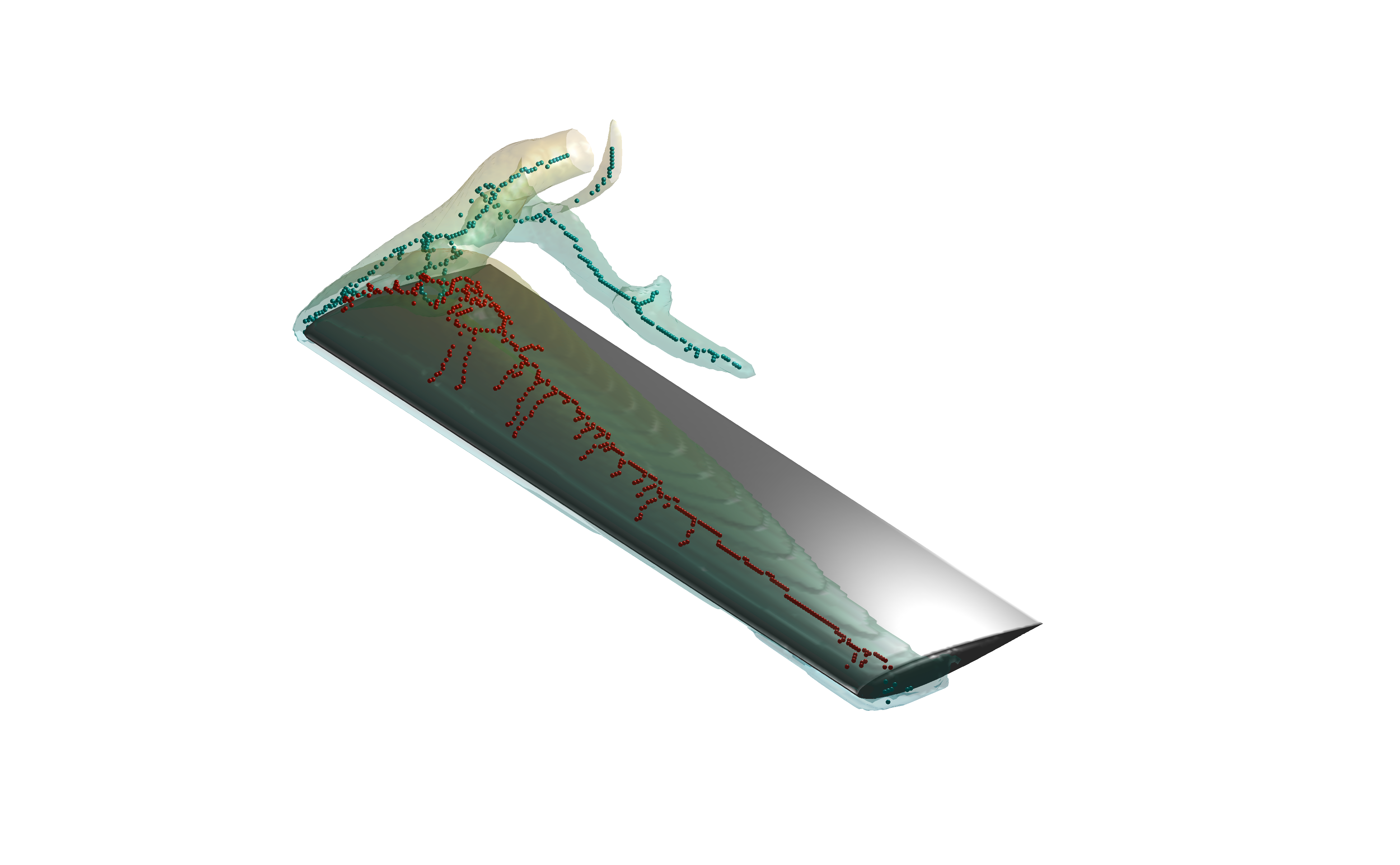}};
   \node(C) at (A.south)[below] {\includegraphics[width=\twhalf,trim=3.2cm 1.25cm 4cm 1.25cm, clip]{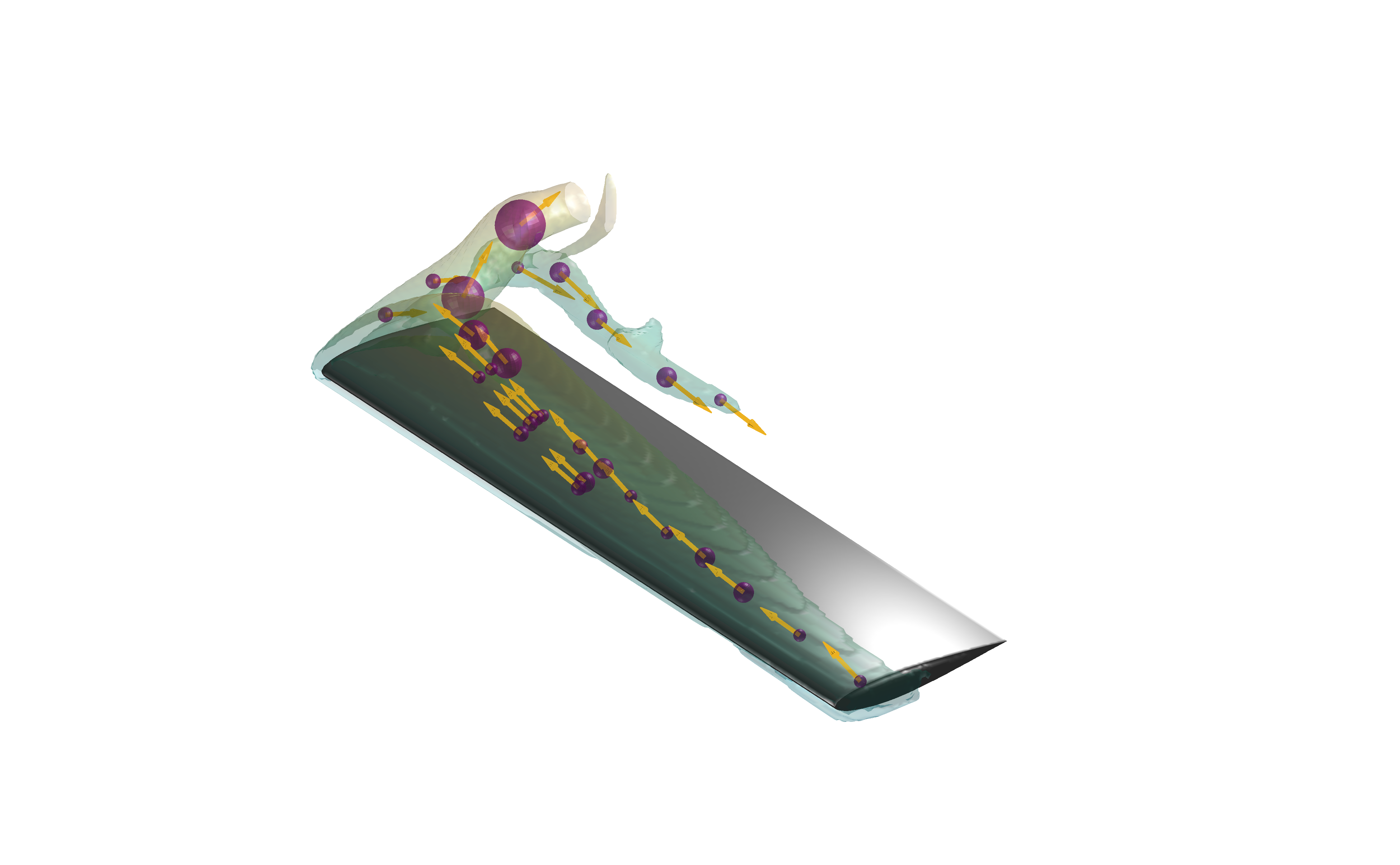}};
   \node(D) at (B.south)[below] {\includegraphics[width=\twhalf,trim=3.2cm 1.25cm 4cm 1.25cm, clip]{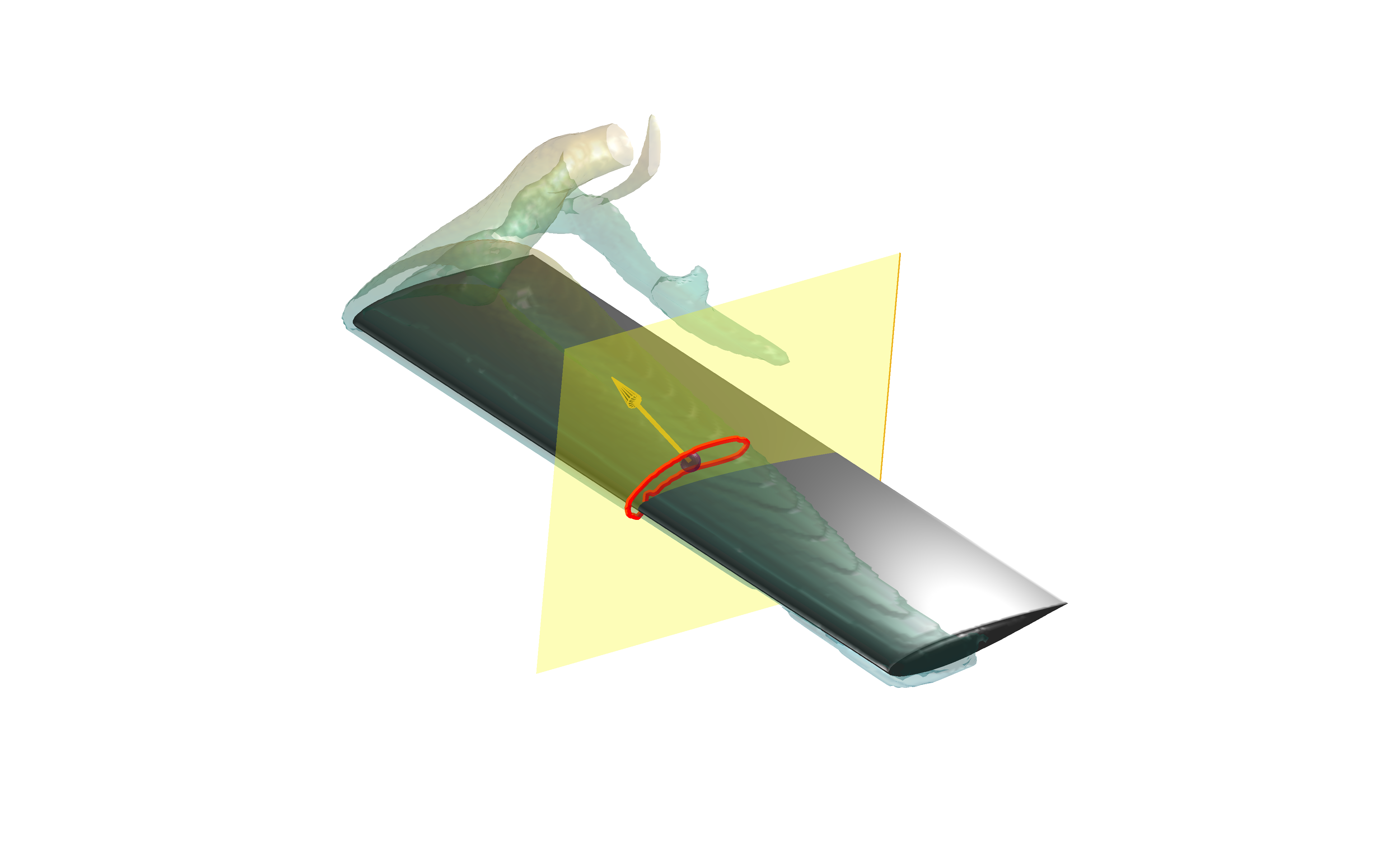}};
   %
   \node(CA) at ([yshift=0.80cm]A.south)[below]{$a)$};
   \node(CB) at ([yshift=0.80cm]B.south)[below]{$b)$};
   \node(CC) at ([yshift=0.80cm]C.south)[below]{$c)$};
   \node(CD) at ([yshift=0.80cm]D.south)[below]{$d)$};
  \end{tikzpicture}
  \caption{$(a)$ Isosurface of $Q^\prime = 4u_\infty^2/c^2$ at mid-downstroke ($t/T = 0.25$).
           $(b)$ Skeleton of the vortical structure given by $Q^\prime = 4u_\infty^2/c^2$ (translucent). 
           Points corresponding to the LEV in red, rest in green.
           $(c)$ Spheres (in magenta) inscribed in the $Q^\prime = 4u_\infty^2/c^2$ isosurface, centered on selected points of the skeleton. The yellow arrows are $\mathbf{n}^k$, the direction of the relative vorticity averaged in the corresponding sphere. 
           $(d)$ Plane (in yellow) perpendicular to $\mathbf{n}^k$, for a particular point in the skeleton. The red contour corresponds to ${\cal C}^k$ for that point of the skeleton. \reva{Isosurfaces of $Q^\prime$ are colored with the vertical coordinate 
           of the LEV (i.e., $z/c$) with color transitioning from green to yellow as the 
           distance from the wing's chord line increases.}
  \label{fig:FVA} }
 \end{center}
\end{figure}

\def\tmpa{FVA_isosurface_relQ}
\def\tmpb{FVA_vortex_graph_relQ}
\def\tmpc{FVA_vortex_graph_relQ}
\def\tmpd{FVA_dirvecs_planes_relQ_paper_figure}
\def\tmpzz{2W-AR4-R000_Qth4000}
\def\tmpz{2W-AR4-R000_Qth4000_ifr213}
\begin{figure}
 \begin{center} 
  \begin{tikzpicture}
   \coordinate(O) at (0.,0.); 
   %
   \node(A) at (O)              {\includegraphics[scale=0.9,trim=3.1cm 1.25cm 3.7cm 1.25cm, clip]{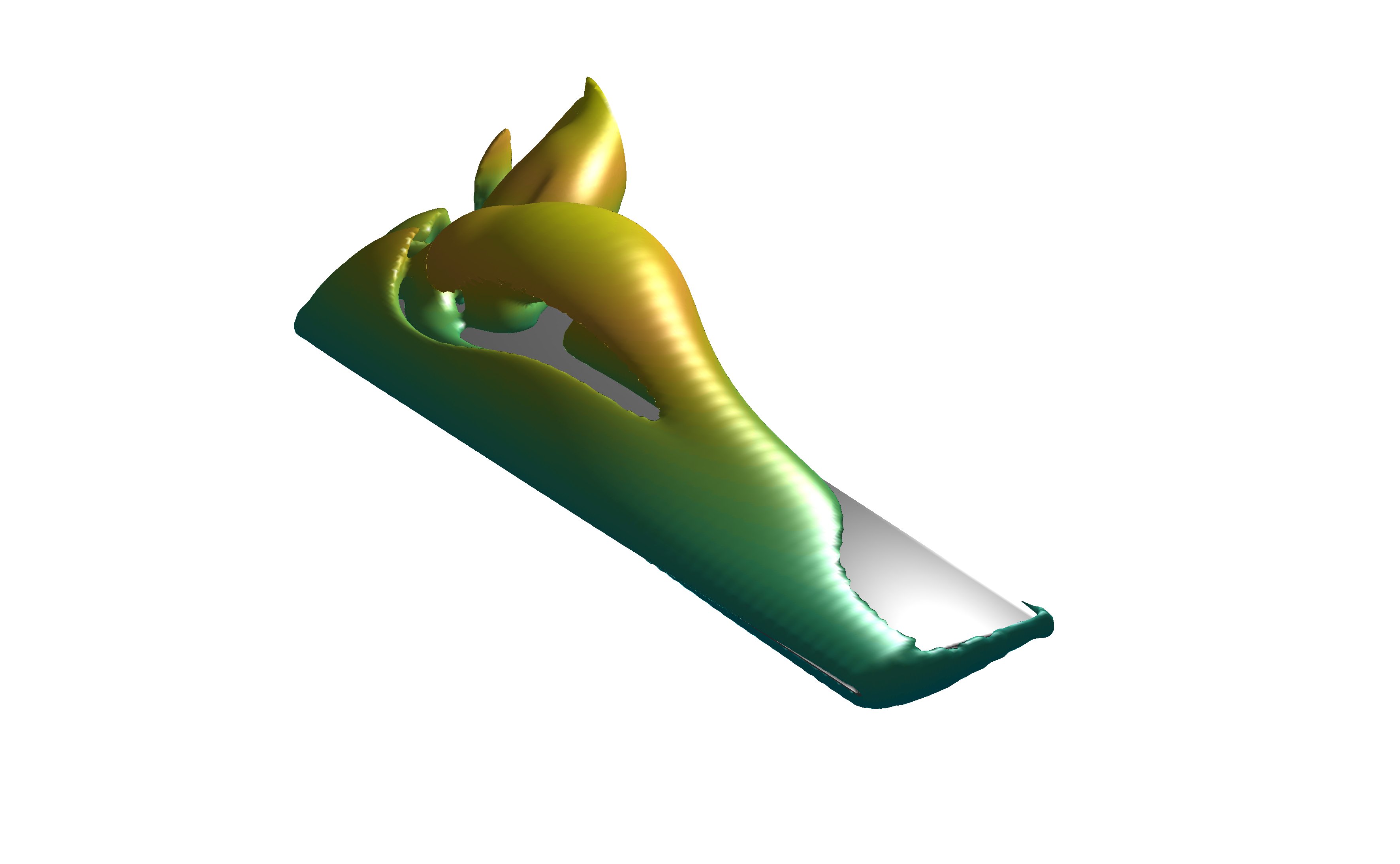}};
   \node(B) at (A.east)[right]  {\includegraphics[scale=0.9,trim=3.1cm 1.25cm 3.7cm 1.25cm, clip]{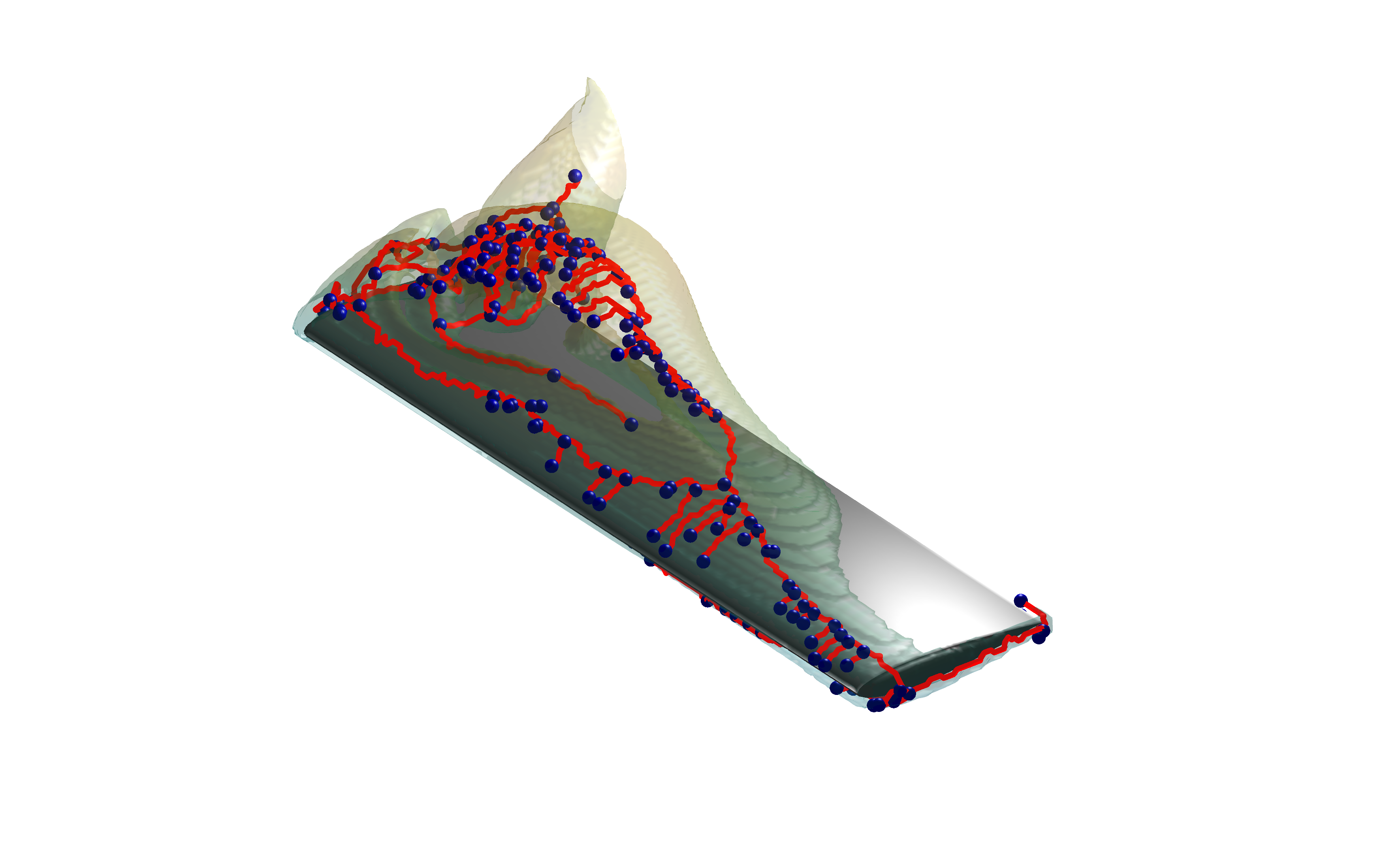}};
   \node(C) at (A.south)[below] {\includegraphics[scale=0.9,trim=3.1cm 1.25cm 3.7cm 1.25cm, clip]{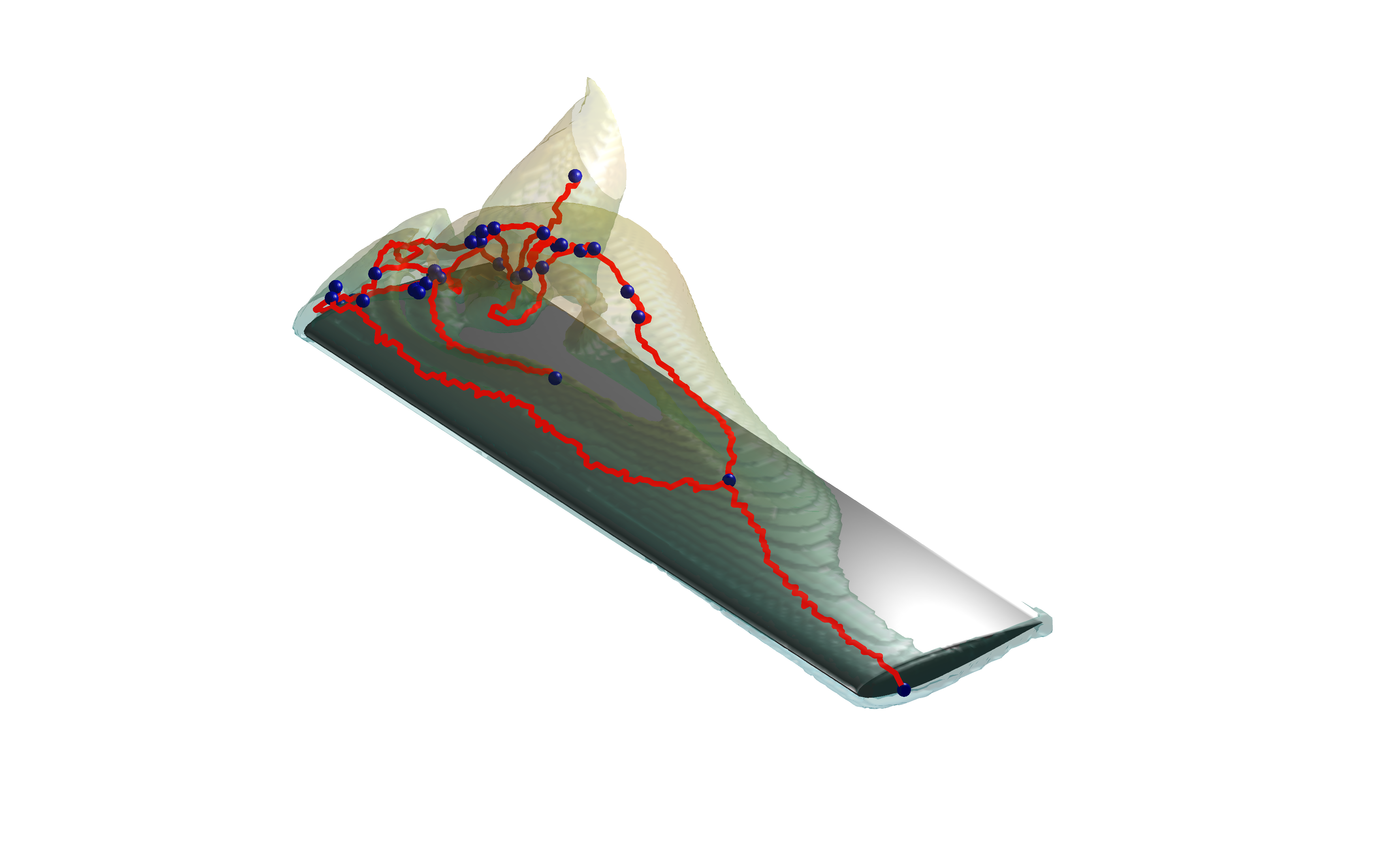}};
   \node(D) at (B.south)[below] {\includegraphics[scale=0.9,trim=3.1cm 1.25cm 3.7cm 1.25cm, clip]{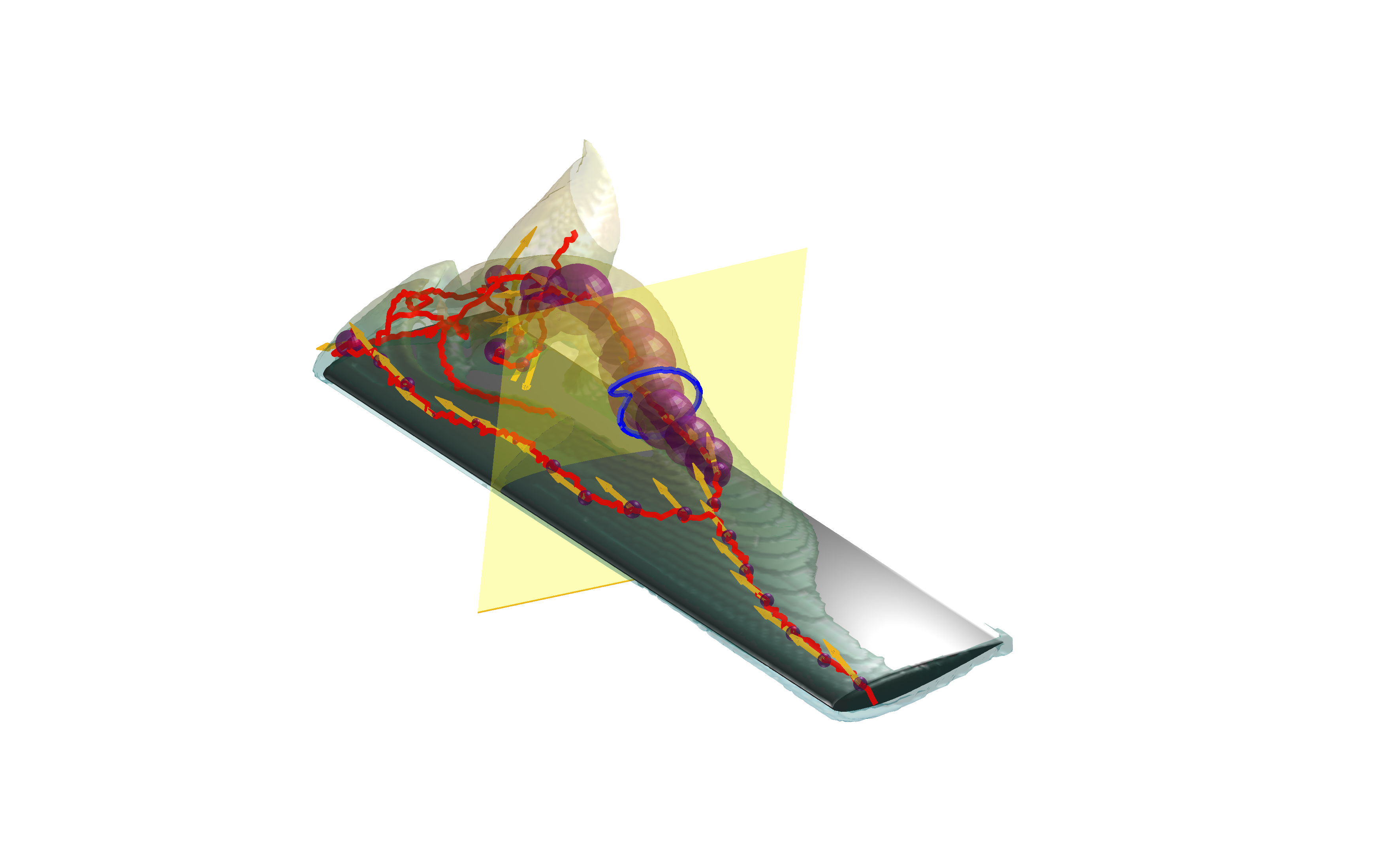}};
   %
   \node(CA) at ([yshift=0.80cm]A.south)[below]{$a)$};
   \node(CB) at ([yshift=0.80cm]B.south)[below]{$b)$};
   \node(CC) at ([yshift=0.80cm]C.south)[below]{$c)$};
   \node(CD) at ([yshift=0.80cm]D.south)[below]{$d)$};
  \end{tikzpicture}
  \caption{
  \reva{
  $(a)$ Isosurface of $Q^\prime = 4u_\infty^2/c^2$ at $t/T = 0.41$.
           $(b)$ Skeleton graph of the vortical structure given by
           $Q^\prime = 4u_\infty^2/c^2$ (translucent).
           Links/branches are shown in red and nodes between them in blue.
           $(c)$ Links/branches and nodes after the discrimination process of
           the non-physical vortex ramifications.
           $(d)$ Spheres (in magenta) inscribed in the 
           $Q^\prime = 4u_\infty^2/c^2$ isosurface, centered on selected points
           of the skeleton.
           The yellow arrows are $\mathbf{n}^k$, the direction of the relative
           vorticity averaged in the corresponding sphere.
           Selected links/branches are depicted in red.
           Plane (in yellow) perpendicular to $\mathbf{n}^k$, for a particular
           point of the skeleton.
           The blue contour inscribed on the plane corresponds to ${\cal C}^k$
           for that point of the skeleton.
           Isosurfaces of $Q^\prime$ are colored with the vertical coordinate 
           of the LEV (i.e., $z/c$) with color transitioning from green to yellow as the 
           distance from the wing's chord line increases.}
  \label{fig:FVA2} }
 \end{center}
\end{figure}

The skeleton provided by the thinning algorithm for the case used as an example in the previous subsection is shown in figure \ref{fig:FVA}$b$ with red and green dots. The resulting skeleton follows reasonably well the overall shape of the vortical structure (i.e., the translucent object), although near the leading edge of the wing the skeleton shows extensive branching. 
The origin of this branching is \reva{linked} to the shape of the vortical structure in 
those locations, which resembles a cylindrical vortex joined to the leading edge by a thin shear layer. 
The chordwise oriented branches develop along this shear layer. This seems to be a spurious result of the 
thinning algorithm, and can be easily reproduced by applying the thinning algorithm to 3D objects 
obtained by joining a slender cylinder with a flat plate. 

\reva{
It is also evident from figure \ref{fig:FVA}$b$ that not all skeleton points  belong to the LEV. 
At the threshold depicted in figure \ref{fig:FVA}, the LEV is connected to the tip vortex (TiV) at the outboard wing tip, as well as to a section of 
a trailing edge vortex (TEV) shed earlier. 
To assess which skeleton points belong to the LEV, we use 
two geometrical criteria based on the position of the skeleton points, and the orientation of the vortical structure at these points.
}

\reva{
Determining the orientation of the vortical structure at each skeleton point  is not straightforward, as the distribution of points 
provided by the thinning algorithm is not smooth and the skeleton contains branches. 
This is visible at the first time instant, Fig. \ref{fig:FVA}b, and even more so
at the second instant, Fig. \ref{fig:FVA2}b.
In  Fig. \ref{fig:FVA2}b, 
the skeleton provided by the thinning algorithm is shown with a network
graph composed by red links or branches and the nodes between then in blue.
}

\reva{
The orientation of the vortical structure is defined in terms of the direction of the local vorticity, averaged within a region surrounding each skeleton point. 
Somewhat arbitrarily, this region is defined as the largest sphere 
inscribed in the isosurface $Q^\prime = Q^\prime_{th}$ 
and centered at each skeleton point. 
These spheres are defined as collections of voxels, and skeleton points for which corresponding sphere contains only one voxel are discarded. 
The volume associated to the $k$-th skeleton point is denoted as $V^k$, and the corresponding direction, based on averaged local vorticity, is denoted as $\mathbf{n}^k$. 
}

\reva{
It is possible to remove non-physical vortex ramifications following a  discrimination process
removing the branches whose direction is not aligned with the direction of the averaged 
local vorticity.
The result of this process can be seen in Fig. \ref{fig:FVA2}c
by comparing to Fig. \ref{fig:FVA2}b.
The spheres and corresponding vectors $\mathbf{n}^k$ are shown, for the first instant considered, in figure \ref{fig:FVA}$c$ for selected points along the skeleton of the vortical structure. 
Interestingly, the points within the branches appearing near the leading edge of the wing exhibit roughly the same local vorticity direction, primarily 
pointing towards the outboard wing tip. 
In contrast, for the segments resembling the TiV and TEV, the local vorticity is primarily chordwise and spanwise (towards the inboard wing tip), respectively. 
For the second instant considered, the spheres and corresponding vectors are shown in 
figure \ref{fig:FVA2}$d$.
}

\reva{
Once the local direction of the skeleton points of the vortical structures is defined and computed, LEV points are determined by the following conditions: 
}
\begin{equation}
\label{eq:orient}
\left[\mathbf{n}^k - (\mathbf{n}^k\cdot \mathbf{e}_{z}) \mathbf{e}_{z} \right] \cdot \mathbf{e}_{y} \le \cos(\theta_{th}), 
\end{equation}
\begin{equation}
\label{eq:zpos}
z_s^k \ge 0, 
\end{equation} 
\reva{
where $z_s^k$ is the vertical coordinate of the $k$-th skeleton point. 
Physically, equation (\ref{eq:orient}) requires that the angle between the spanwise direction of the wing (i.e., $\mathbf{e}_{y}$) and the projection of $\mathbf{n}^k$ onto the $x-y$ plane is smaller than a threshold angle, $\theta_{th}$.  
For moderate to small values of $\theta_{th}$, this is equivalent to requiring 
that the direction of the vortex skeleton is more or less aligned with the spanwise direction. 
Note that this condition stems from the 
rectangular shape of our wings.
For  wings with different geometries, it may be more appropriate to consider the angle between
$\mathbf{n}^k$ and a local direction parallel to the leading edge 
of the wing (i.e., with a varying $\theta_{th}$ along $y$). 
Finally, equation (\ref{eq:zpos}) discriminates points of the skeleton in the lower surface of the wing, since we are analyzing solely the downstroke. 
}

Figure \ref{fig:FVA}$b$ \reva{illustrates} the result of applying these geometrical conditions \reva{to}  the \reva{vortex skeleton points} \reva{at} the mid-downstroke. 
%
%
\reva{Skeleton} points satisfying equations (\ref{eq:orient}) and (\ref{eq:zpos}), with $\theta_{th}=30^\circ$, are colored in red and they correspond to the section of the vortical structure that is easily identified with the LEV. Points failing to satisfy all conditions are colored in green, and they correspond to the TiV and TEV.   
It should be noted that, although not shown here, several values of $\theta_{th}$
have been tested in the present case. The observed differences were negligible when $25^\circ \le \theta_{th} \le 60^\circ$, except (maybe) at the end of the downstroke, when the displacement of the LEV is maximum.  

\subsection{Computing averaged quantities along the LEV}

The last step of the method is to evaluate flow variables along the LEV.
This is done with a procedure analogous to that used in previous works
\citep{Jones2011,jardin2014,calderon2014,arranz2018b}.
At each point of the skeleton belonging to the LEV, 
a plane perpendicular to $\mathbf{n}^k$ is defined. 
The intersection of that plane with the volume 
satisfying $Q^\prime > Q^\prime_{th}$ is denoted 
${\cal C}^k$ (shown in figures \ref{fig:FVA}$d$ \reva{and 
\ref{fig:FVA2}$d$}). 
Any physical variable of interest, $\phi$, is averaged over  
${\cal C}^k$ to provide $\phi_s^k$. 
This applies to the velocity, vorticity and pressure. 
The local circulation in this plane is defined as 
\begin{equation}
\Gamma_s^k = \int_{{\cal C}^k}\boldsymbol{\omega}^\prime\cdot\mathrm{d}\boldsymbol{S},
\end{equation}
where $\mathrm{d}\boldsymbol{S}$ is the differential element of surface.
Note that in previous works
the chosen plane is a chordwise-vertical plane, which assumes a LEV aligned with the spanwise direction \citep{Jones2011,jardin2014,calderon2014,arranz2018b}. 
The present choice of plane is more general, allowing for a deformed LEV, reasonably aligned (i.e., see $\theta_{th}$ in equation \ref{eq:orient}) with the leading edge of the wing.  

It should be noted that the objective of the method presented here is to provide a quantitative description of the LEV along its core.
In the present case, the core is roughly aligned along the spanwise direction. 
Hence, the positions and physical quantities on the skeleton of the LEV, 
$(x_s^k,y_s^k,z_s^k)$ and $\phi_s^k$, are averaged in spanwise bins, 
to characterize the LEV as a function of the spanwise coordinate $y$ and time.  
The position of the LEV core in a spanwise bin of width $\Delta$ (i.e., $y\pm\Delta/2$)  is
given by the point of the skeleton with the largest sphere volume, $V_{\max}(y)=\max(V^k)$.
Since this volume is computed as a sum of voxels, it is possible to find several points
within a bin with the same $V^k$. 
Hence, formally, the position of the LEV core, $\mathbf{x}_c$, is  
defined as the averaged position of the points of the skeleton 
inside the bin whose $V^k$ is equal to the maximum $V^k$ on the bin. 
Mathematically, 
\begin{equation}
\label{eq:x_c}
\mathbf{x}_c(y) = \frac{1}{N_k}\sum_k \mathbf{x}_s^k 
\mbox{,~~for~} k \mbox{~such that~} y_s^k\in[y\pm\Delta/2] \mbox{~and~} V^k = V_{\max}(y), 
\end{equation}
where $N_k$ is the number of points in the skeleton satisfying the condition in equation (\ref{eq:x_c}). 
The same average is used to define physical quantities along the core (i.e., pressure, velocity, vorticity and local circulation), 
\begin{equation}
\label{eq:phi_c}
\phi_c(y) = \frac{1}{N_k}\sum_k \phi_s^k  
\mbox{,~~for~} k \mbox{~such that~} y_s^k\in[y\pm\Delta/2] \mbox{~and~} V^k = V_{\max}(y). 
\end{equation}

Note that the definitions of $\mathbf{x}_c$ and $\phi_c$ in equations (\ref{eq:x_c}) and (\ref{eq:phi_c}) are explicitly designed to deal with the branching appearing in  \ref{fig:FVA}.
The points in the branches usually have smaller spheres (i.e., smaller $V^k$), as the vortical structure around them is thinner. 
Since the core of the LEV is expected to be associated to the thicker 
region of the vortical structure, the definition of the position of 
the LEV core ($\mathbf{x}_c$ in equation \ref{eq:x_c}) and the physical
quantities inside it ($\phi_c$ in equation \ref{eq:phi_c}) only considers 
the largest spheres in the bin. 
However, the points along the chordwise branches have essentially 
the same orientation as the point at the intersections, as observed 
in figure \ref{fig:FVA}$c$. 
Hence, $\phi_c$ defined in equation 
(\ref{eq:phi_c}) is virtually indistinguishable from a standard or 
volume-weighed average over all the points of the skeleton in the bin, 
due to the little variation of $\mathbf{n}^k$ along the branches. 

%
\section{Results}
\label{sec:results}
The identification method has been applied to the two cases described in section \ref{sec:cases}. 
For both cases, the LEV skeleton is obtained at various instants during the 
downstroke using $\theta_{th} = 30^\circ$. 
Several values of the threshold $Q^\prime_{th}$ have been used, 
to asses the effect that the threshold has on the characterization of the LEV. 
Finally, positions and physical variables along the vortex core are 
computed using equations (\ref{eq:x_c}) and (\ref{eq:phi_c}) 
with spanwise bins of width $\Delta = 4 h$, where $h=c/56$ is the grid spacing of the simulation. 
The uncertainty in the position of the LEV is measured with the 
maximum and minimum coordinates of all the points of the skeleton within a bin.
The uncertainty in the physical variables ($\phi_c$) is computed as the 
standard deviation of $\phi_s^k$ for all skeleton points inside two consecutive 
bins with respect to the mean value of $\phi_c$ in these two bins. 
These uncertainties are shown with shaded contours in the figures below. 

Figure \ref{fig:LEVskpos_tT250} shows the streamwise ($x_c$) and vertical ($z_c$)
position of the LEV core as a function of the spanwise coordinate at a fixed time instant, 
$t/T = 0.25$ (mid-downstroke).
At that time instant the LEV is already developed.
In fact, although not shown here, a peak of lift appears slightly before the mid-downstroke  \citep{gonzalo:2018}. 
Panels $a$ and $b$ of figure \ref{fig:LEVskpos_tT250} show that, for both cases, 
the LEV separates vertically from the wing close to the outboard wing tip.
Panels $c$ and $d$ of figure \ref{fig:LEVskpos_tT250} show that the LEV core is found further
downstream when increasing the spanwise coordinate, except very close to the wing tip. 
The uncertainty in the position of the LEV core is small except for
$x_c$ in the region where the LEV is farther away from the wing and 
branching of the LEV skeleton is more apparent (see red points in figure \ref{fig:FVA}$b$).
Comparing both cases, figures \ref{fig:LEVskpos_tT250}$a$ and $b$ show that 
the maximum height of the LEV core seems to be independent of $AR$.
This might be related to the design of the cases, both having the same 
vertical displacement of the outer wing tip. 
With respect to the streamwise position of the LEV core, it seems that 
there is indeed a non-negligible difference between cases of $AR=2$ 
and $AR=4$, panels $c$ and $d$ of figure \ref{fig:LEVskpos_tT250}.
However, this difference is difficult to quantify since it is of the same order as the uncertainty.
%

\def\tmpa{LEV_sk_pos_yw_th_influence_relQ_noxth}
\def\tmpb{2W-AR2-R000_tT250}
\def\tmpc{2W-AR4-R000_tT250}
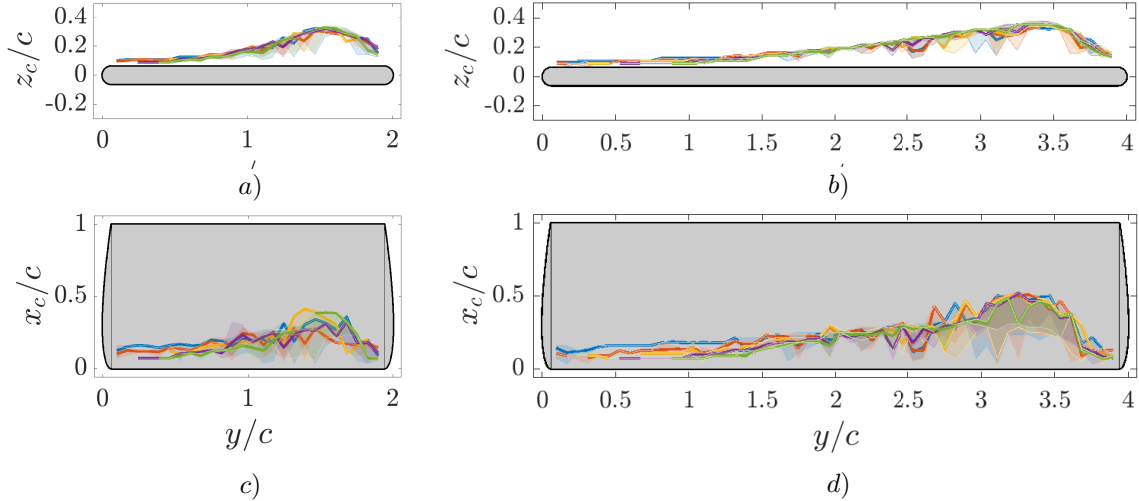
\begin{figure}
 \begin{center} 
  \begin{tikzpicture}
   \coordinate(O) at (0.,0.); 
   \setkeys{Gin}{width=0.3687\linewidth}
   \node(A) at (O) {\includegraphics[scale=0.5,trim=0cm 0cm 0cm 0cm, clip]{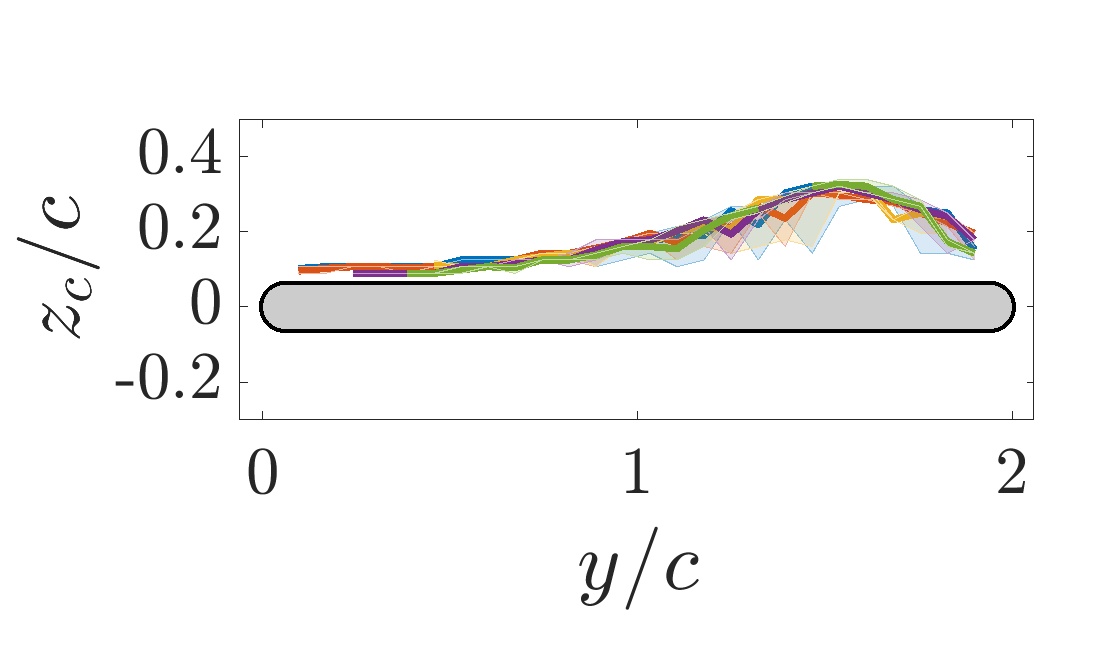}};
   \node(C) at ([yshift=0.82cm]A.south)[below] {\includegraphics[scale=0.5,trim=0cm 0cm 0cm 0cm, clip]{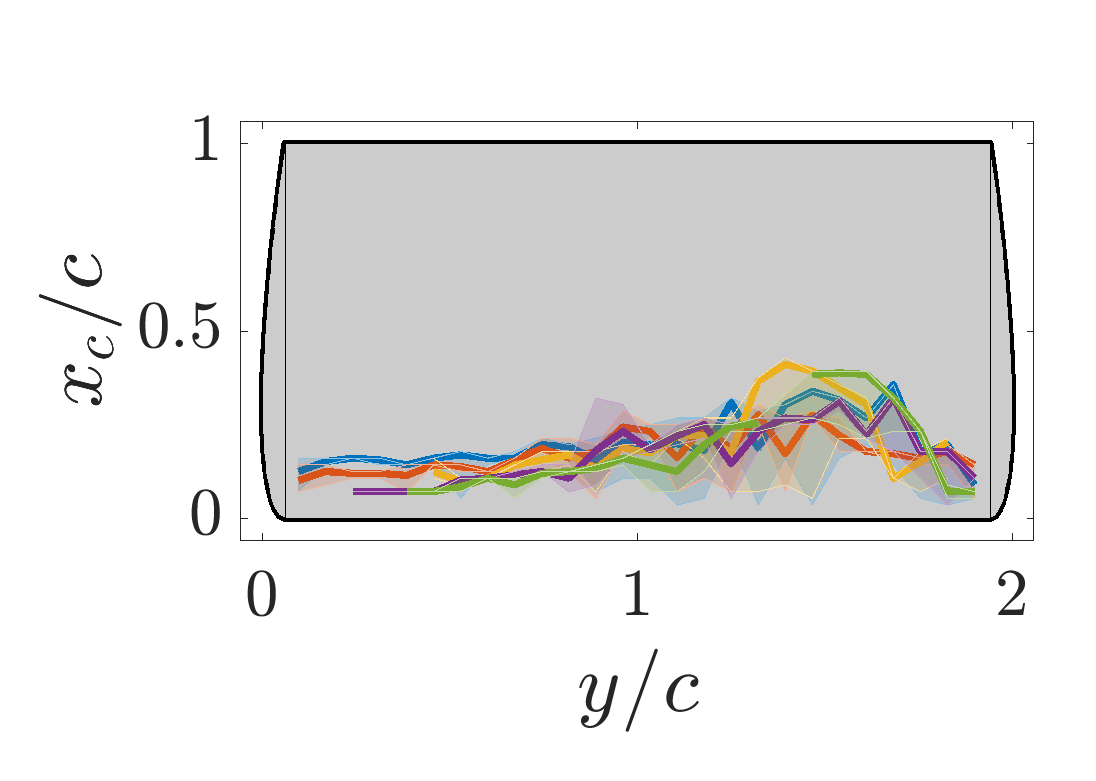}};
   \setkeys{Gin}{width=0.6250\linewidth}
   \node(B) at (A.east)[right] {\includegraphics[scale=0.5,trim=0.00cm 0cm 0cm 0cm, clip]{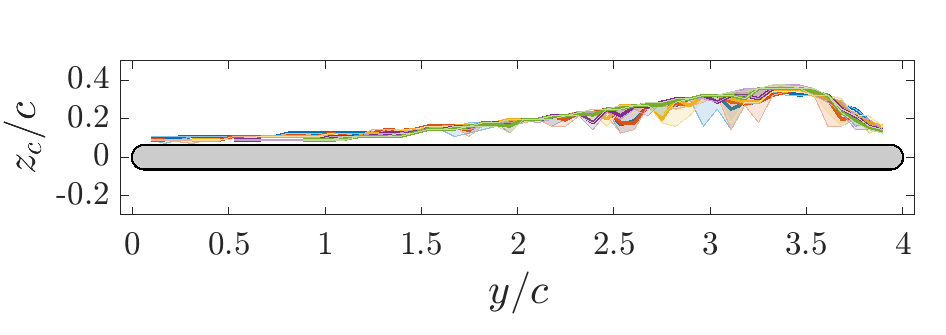}};
   \node(D) at ([yshift=0.02cm]C.east)[right] {\includegraphics[scale=0.5,trim=0.00cm 0cm 0cm 0cm, clip]{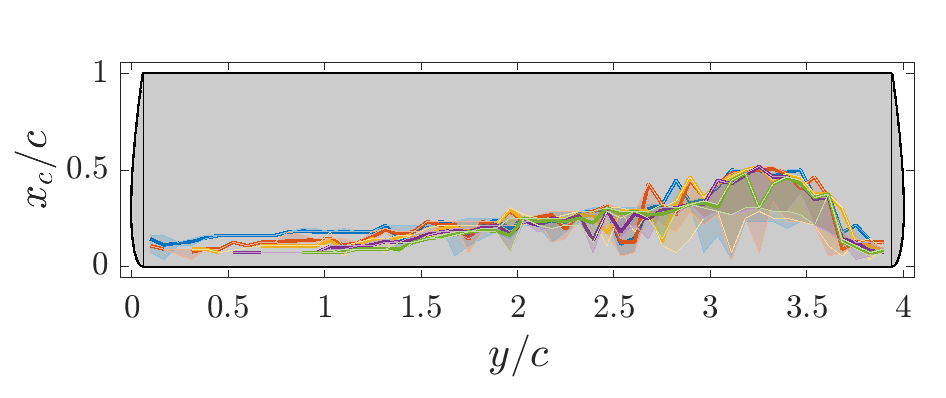}};
   %
   \node(CA) at ([xshift=0.50cm,yshift=0.75cm]A.south)[below]{$a)$};
   \node(CB) at ([xshift=0.50cm,yshift=0.75cm]B.south)[below]{$b)$};
   \node(CC) at ([xshift=0.50cm,yshift=0.20cm]C.south)[below]{$c)$};
   \node(CD) at ([xshift=0.50cm,yshift=0.20cm]D.south)[below]{$d)$};
  \end{tikzpicture}
  \caption{$(a,b)$ Vertical and $(c,d)$ streamwise coordinate
           of the LEV core along the wing span at mid-downstroke
           ($t/T=0.250$). 
           $(a)$ and $(c)$ correspond to $AR = 2$, 
           $(b)$ and $(d)$ to $AR = 4$.
           Lines correspond to 
           $Q^\prime_{th}c^2/u_\infty^2 = 4$ (\solid{c01}),
           $Q^\prime_{th}c^2/u_\infty^2 = 6$ (\solid{c02}),
           $Q^\prime_{th}c^2/u_\infty^2 = 8$ (\solid{c03}),
           $Q^\prime_{th}c^2/u_\infty^2 = 10$ (\solid{c04}) and
           $Q^\prime_{th}c^2/u_\infty^2 = 12$ (\solid{c05}).
           The colored shaded area indicates the uncertainty in the position of the LEV.
           The wing is displayed in grey.
  \label{fig:LEVskpos_tT250} }
 \end{center}
\end{figure}

Next, the influence of the threshold in the position of the LEV core is assessed. 
Overall, the agreement observed in figure \ref{fig:LEVskpos_tT250} for the various thresholds is good.
Note that $x_c(y)$ and $z_c(y)$ are rather irregular. 
However, the observed irregularities do not correspond to a drift when varying the threshold. 
The amplitude of these irregularities seems to be larger for $x_c$ than for $z_c$.
This might be related to the shape of the LEV at this time instant (figure \ref{fig:FVA}$d$),
which is thin along the vertical direction.
Hence, the variations found in the streamwise position of the LEV core are not
translated into its vertical position. 
It is also worth noting that, increasing the threshold, $Q^\prime_{th}$, 
leads to smaller vortical structures.
In the present case, this happens more clearly near the inboard wing tip, 
where the LEV is less intense. 
As a consequence, the LEV and the corresponding lines in figure \ref{fig:LEVskpos_tT250}
become shorter in the spanwise direction with increasing threshold.
Note also that the identification of a smaller LEV in this regions 
results in the LEV core appearing closer to the leading edge 
(i.e., near the inboard wing tip, $x_c \to 0$ as $Q^\prime_{th}$ increases). 

\def\tmpa{LEV_stats_yw_th_influence_relQ_noxth}
\def\tmpb{2W-AR4-R000_tT250}
\begin{figure}
 \begin{center} 
  \begin{tikzpicture}
   \coordinate(O) at (0.,0.); 
   %
   \node(A) at (O) {\includegraphics[width=\twhalf,trim=0cm 0cm 0cm 0cm, clip]{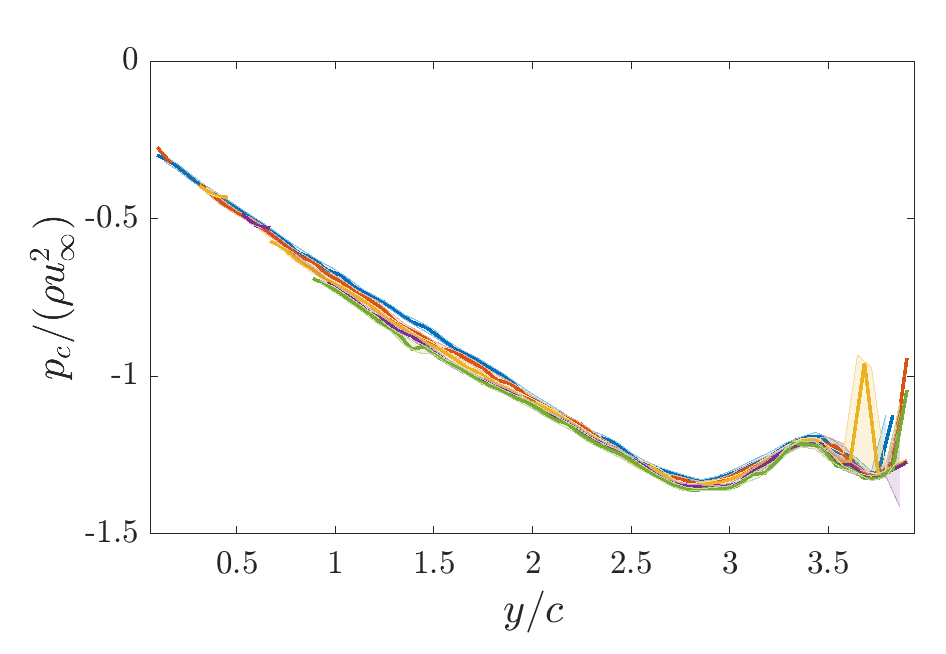}};
   \node(B) at ([xshift=-0.42cm]A.east)[right] {\includegraphics[width=\twhalf,trim=0cm 0cm 0cm 0cm, clip]{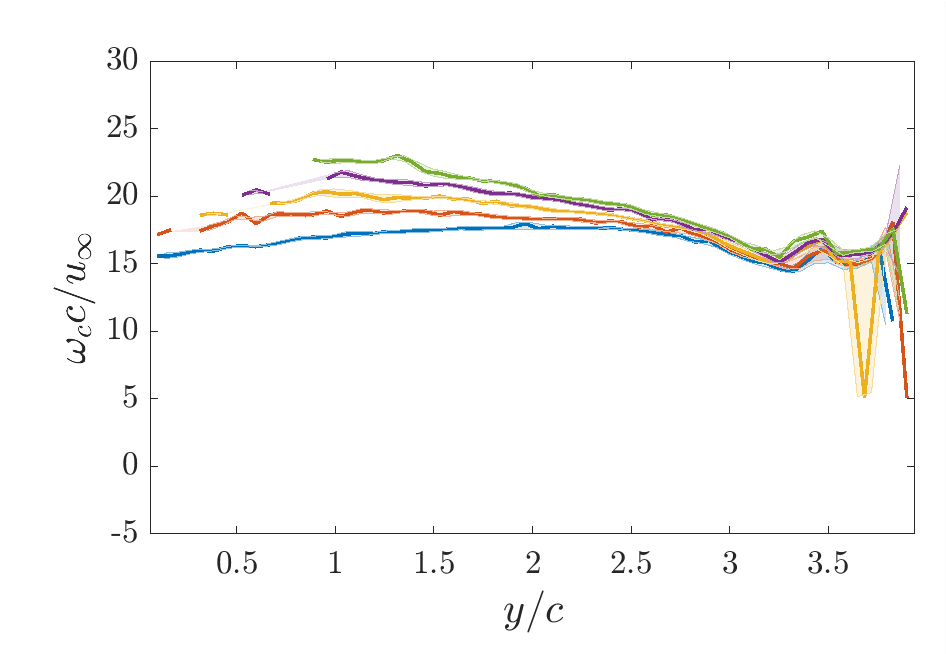}};
   \node(C) at ([yshift=0.67cm]A.south)[below] {\includegraphics[width=\twhalf,trim=0cm 0cm 0cm 0cm, clip]{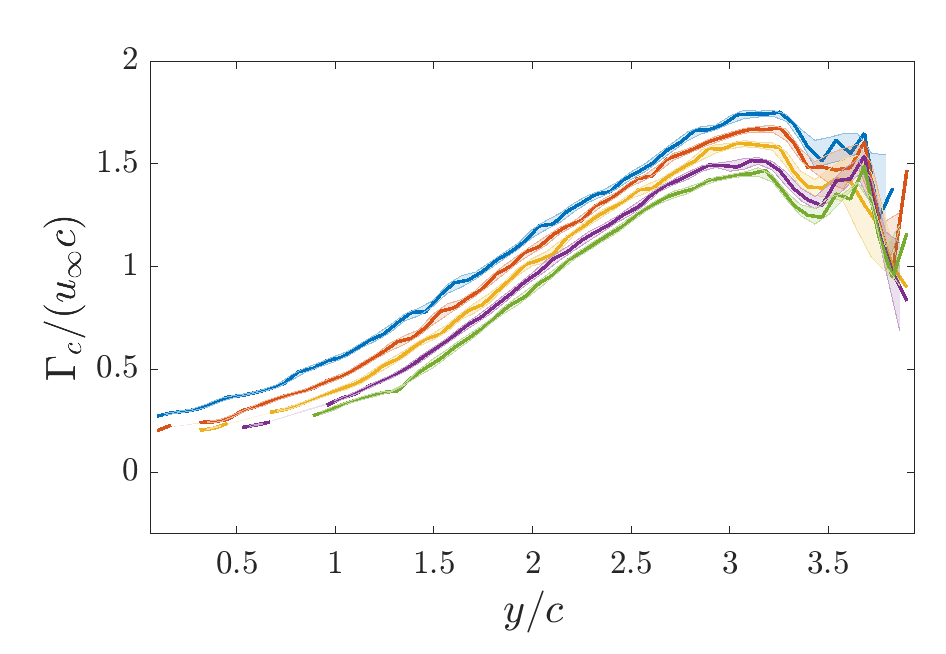}};
   \node(D) at ([yshift=0.67cm]B.south)[below] {\includegraphics[width=\twhalf,trim=0cm 0cm 0cm 0cm, clip]{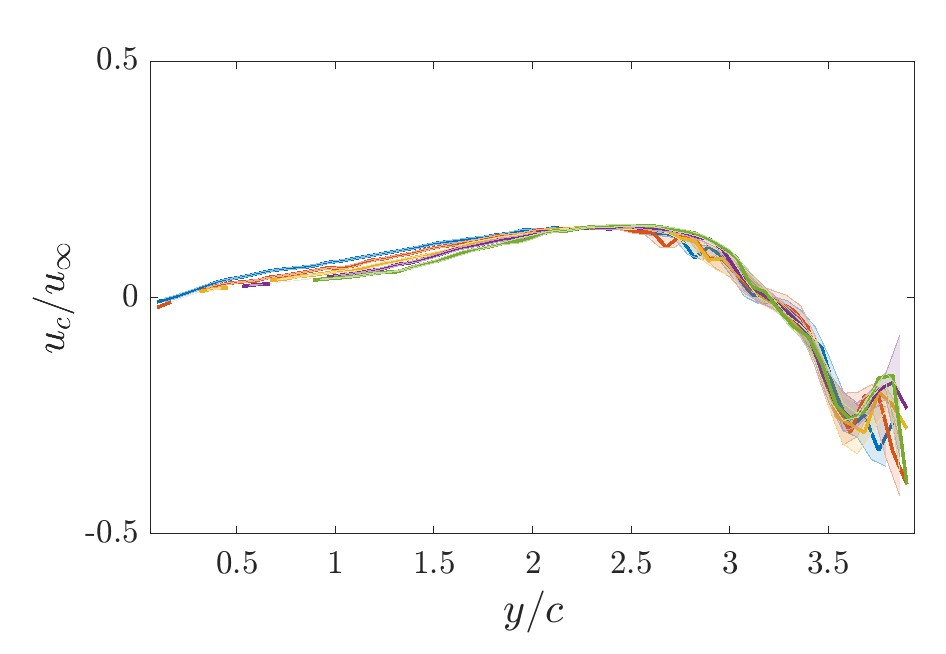}};
   %
   \node(CA) at ([xshift=0.50cm,yshift=0.65cm]A.south)[below]{$a)$};
   \node(CB) at ([xshift=0.50cm,yshift=0.65cm]B.south)[below]{$b)$};
   \node(CC) at ([xshift=0.50cm,yshift=0.20cm]C.south)[below]{$c)$};
   \node(CD) at ([xshift=0.50cm,yshift=0.20cm]D.south)[below]{$d)$};
  \end{tikzpicture}
    \caption{$(a)$ Pressure inside the LEV. $(b)$
           Vorticity along the LEV core. 
           $(c)$ Local circulation.
           $(d)$ Relative velocity along the LEV core.
           All quantities are shown as a function of $y$ 
           and are evaluated at 
           mid-downstroke ($t/T=0.250$). 
           Lines as in figure \ref{fig:LEVskpos_tT250}.
           The colored shaded area indicates the corresponding uncertainty.
  \label{fig:LEVskstats_tT250} }
 \end{center}
\end{figure}
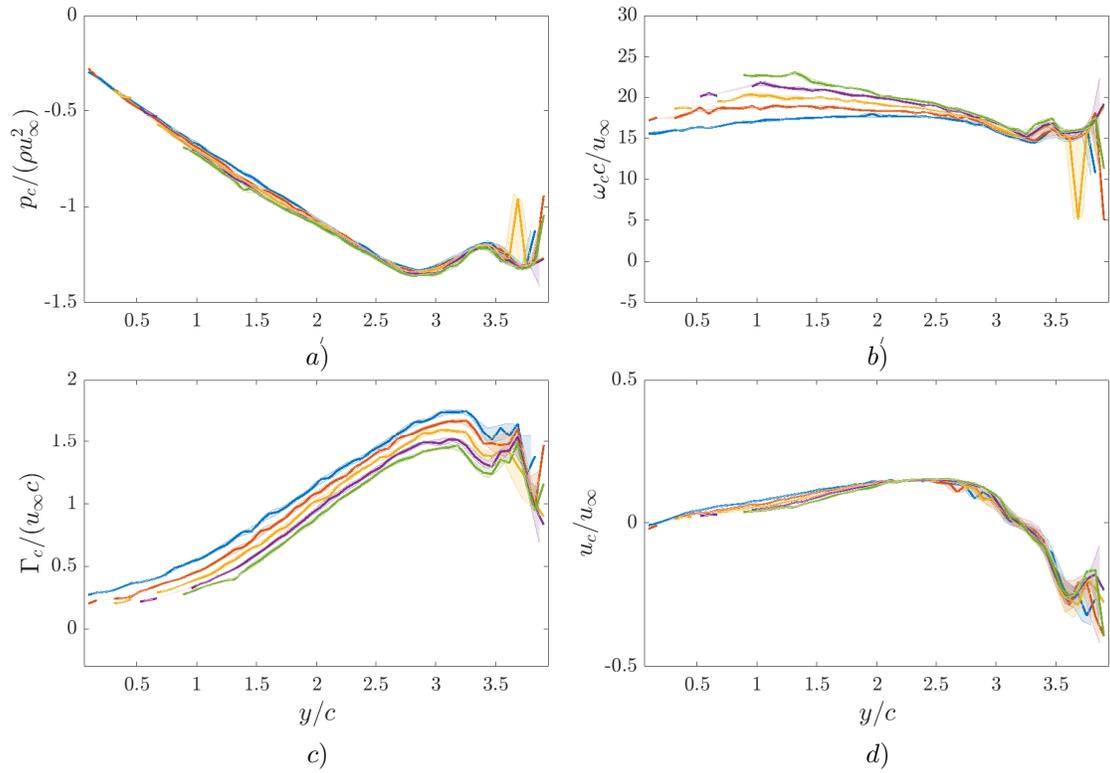

In order to characterize the LEV, some relevant flow quantities are analyzed along
the LEV core. 
The variables considered here are the pressure, $p_c$, the local circulation, $\Gamma_c$, and 
the vorticity and velocity components along the LEV core, $\omega_c=\boldsymbol{\omega}^\prime_c\cdot\mathbf{n}_c$ and $u_c=\boldsymbol{u}^\prime_c\cdot\mathbf{n}_c$, respectively.
These variables are of interest in the LEV dynamics, as shown by previous works \citep{birch2004,jardin2014,jardin2017,arranz2018b}. 
Figure \ref{fig:LEVskstats_tT250} shows the results for the case of $AR=4$. 
As in the case of figure \ref{fig:LEVskpos_tT250}, profiles corresponding to various thresholds are displayed.
Figure \ref{fig:LEVskstats_tT250}$a$ shows that $p_c$ is minimum close to the outboard wing tip.
Roughly at the same location, $\Gamma_c$ is maximum (figure \ref{fig:LEVskstats_tT250}$c$).
On the other hand, the axial vorticity (figure \ref{fig:LEVskstats_tT250}$b$) 
is more uniform, specially for the lower thresholds considered in the figure. 
Finally, the axial velocity (\ref{fig:LEVskstats_tT250}$d$) shows an outboard flow over most of the wing. 
In the region close to the outboard wing tip,  the effect of the 
wing tip vortex yields an inboard flow (i.e., negative $u_c$). 

In terms of the effect of the threshold, figure \ref{fig:LEVskstats_tT250} 
suggests that its effect is somewhat limited in pressure, velocity and circulation. This is more true for pressure and axial velocity than for the local circulation, since the latter is the result of an integral over an area that increases with $Q^\prime_{th}$. 
Not surprisingly, the strongest dependency with the threshold is observed in the axial vorticity: 
increasing $Q^\prime_{th}$ results in a stronger LEV, and consequently the axial vorticity of the LEV increases. 
This dependency is more acute near the inboard wingtip, which suggests that 
the distribution of vorticity within the LEV is more uniform 
near the outboard wing tip. 
Finally, the uncertainty in pressure, local circulation, axial velocity and vorticity is small for all thresholds, except maybe near the outboard wing tip. 

\def\tmpa{LEV_sk_pos_yw_tT_th_influence_relQ_noxth}
\def\tmpb{2W-AR4-R000}
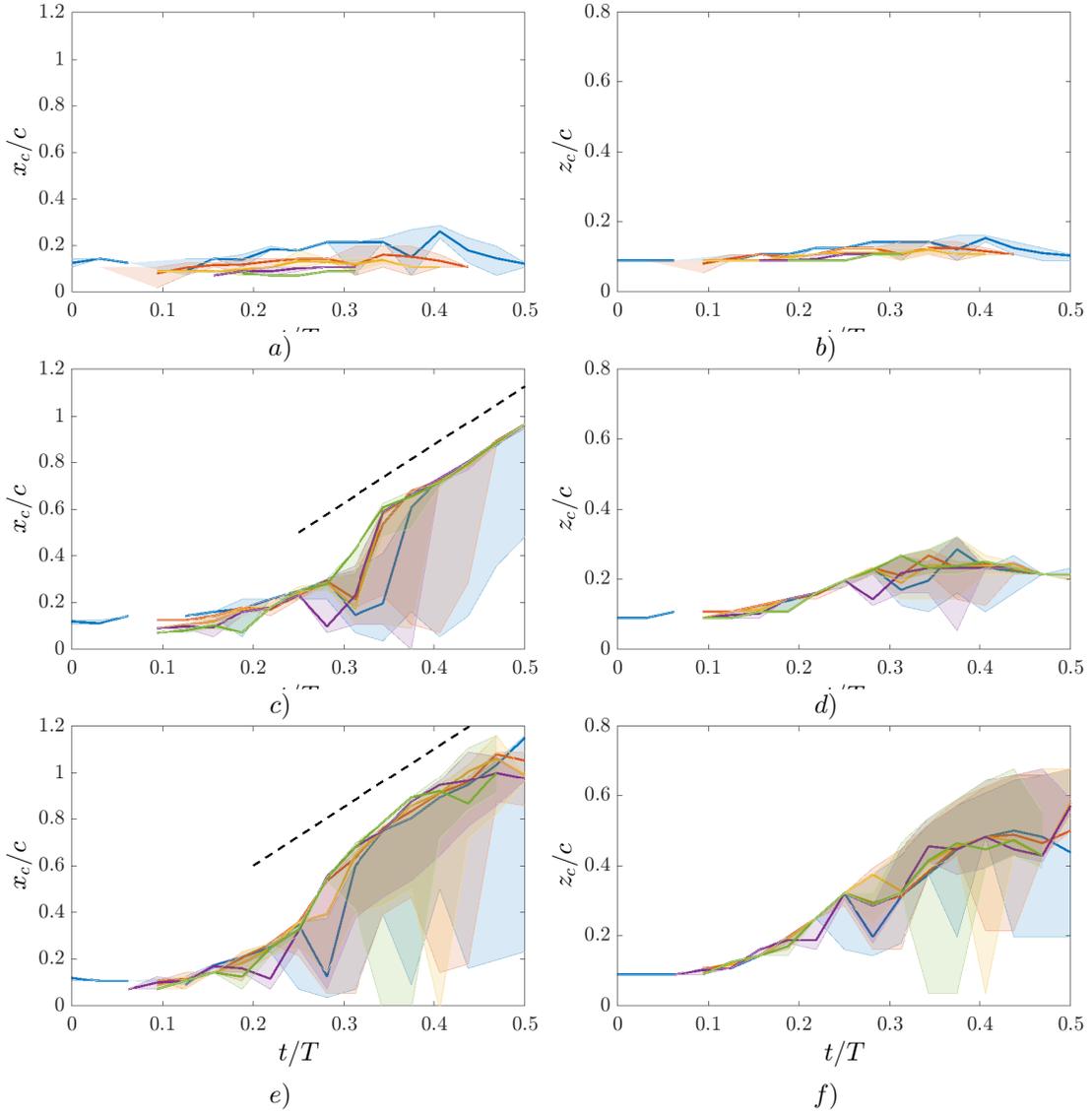
\begin{figure}
 \begin{center} 
  \begin{tikzpicture}
   \coordinate(O) at (0.,0.); 
   %
   \node(A) at (O) {\includegraphics[width=\twhalf,trim=0cm 0cm 0cm 0cm, clip]{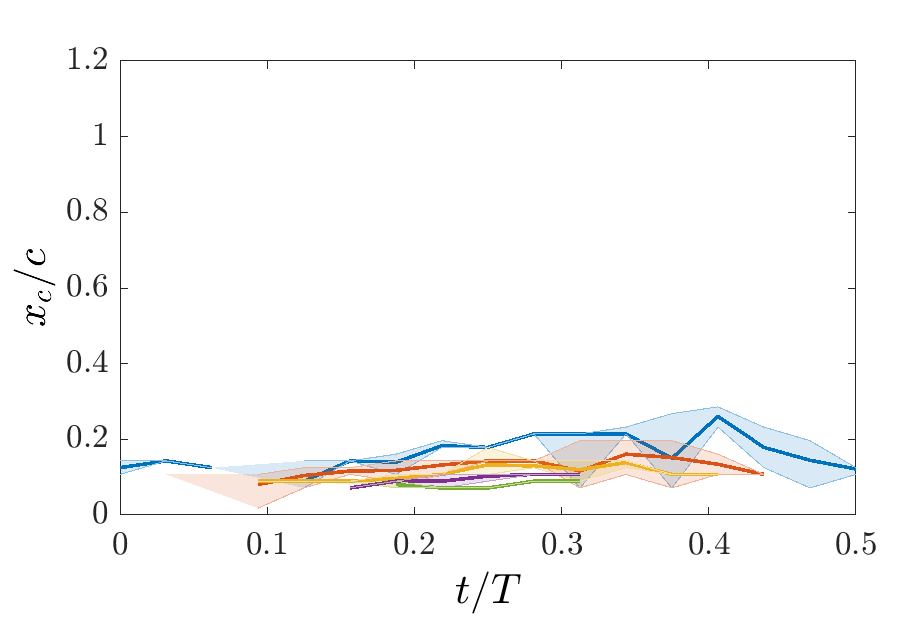}};
   \node(B) at ([xshift=-0.50cm]A.east)[right] {\includegraphics[width=\twhalf,trim=0cm 0cm 0cm 0cm, clip]{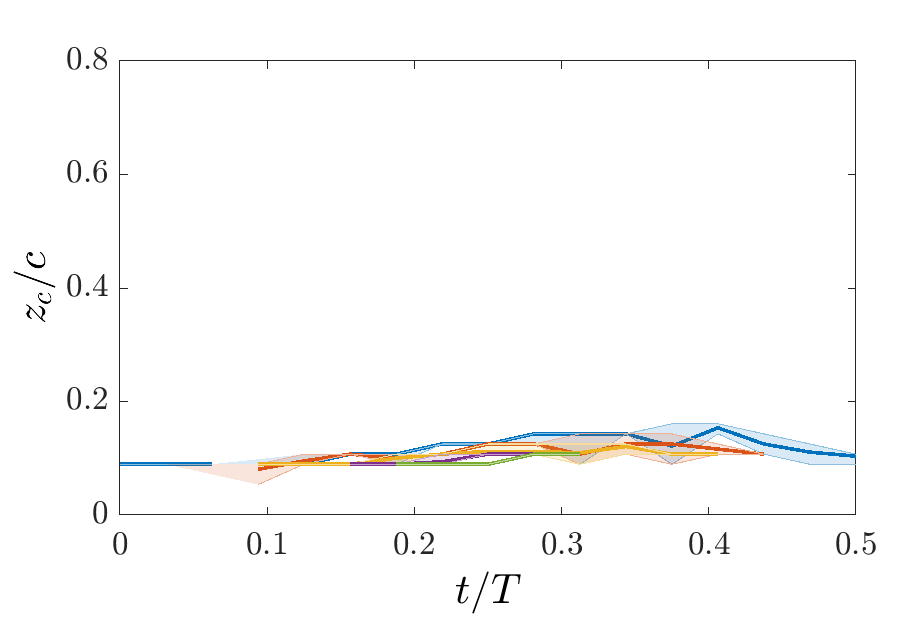}};
   \node(C) at ([yshift=0.71cm]A.south)[below] {\includegraphics[width=\twhalf,trim=0cm 0cm 0cm 0cm, clip]{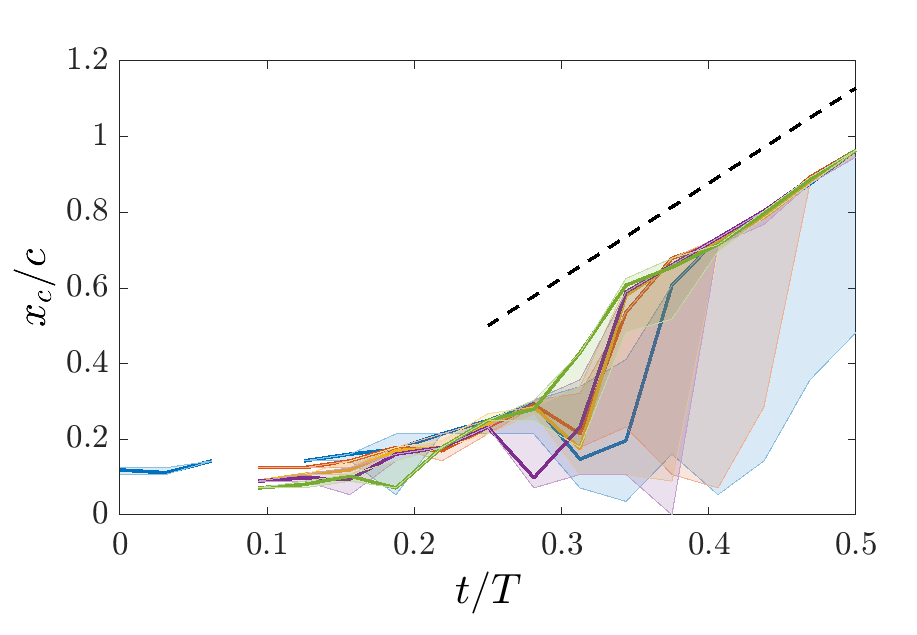}};
   \node(D) at ([yshift=0.71cm]B.south)[below] {\includegraphics[width=\twhalf,trim=0cm 0cm 0cm 0cm, clip]{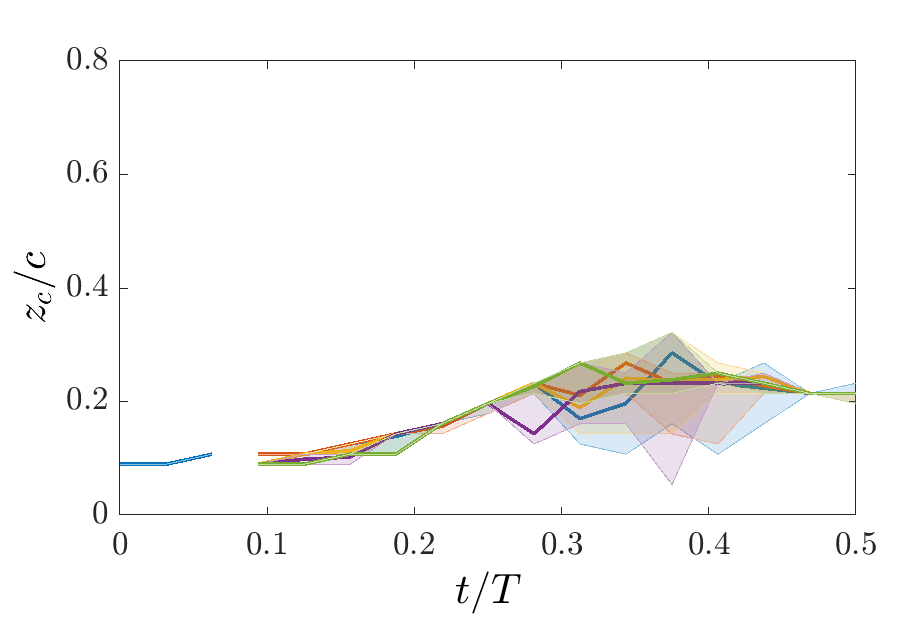}};
   \node(E) at ([yshift=0.71cm]C.south)[below] {\includegraphics[width=\twhalf,trim=0cm 0cm 0cm 0cm, clip]{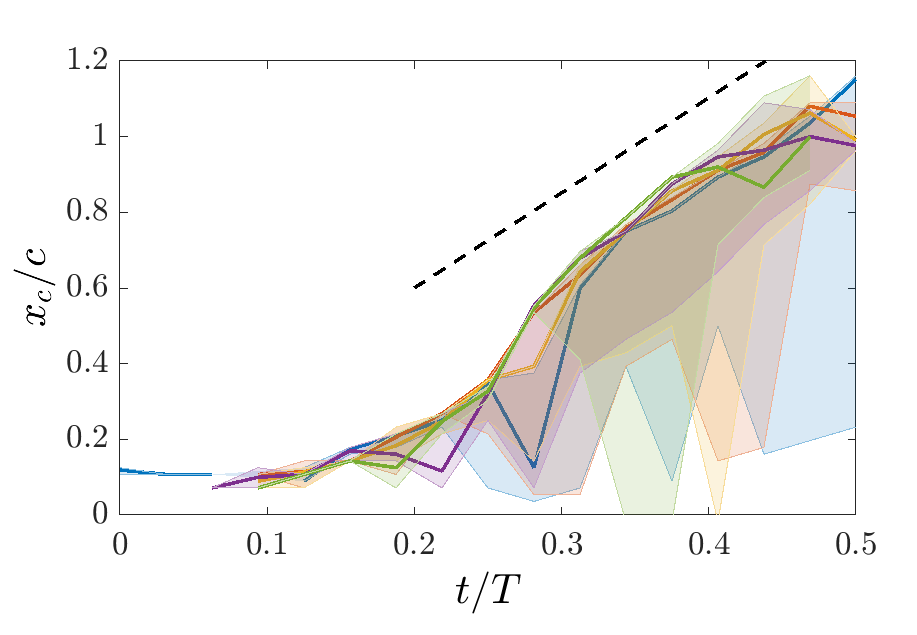}};
   \node(F) at ([yshift=0.71cm]D.south)[below] {\includegraphics[width=\twhalf,trim=0cm 0cm 0cm 0cm, clip]{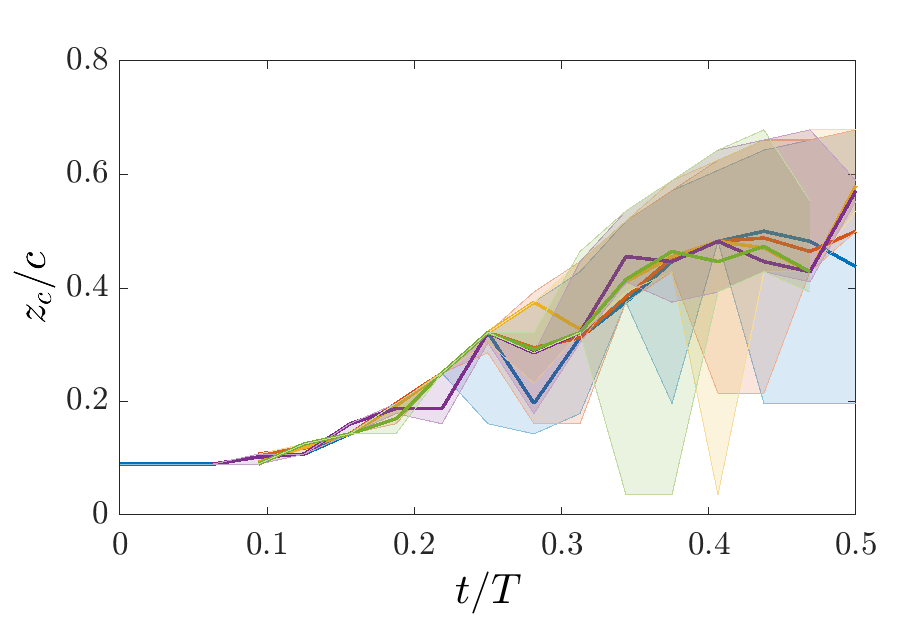}};
   %
   %
   \node(CA) at ([yshift=0.65cm]A.south)[below]{$a)$};
   \node(CB) at ([yshift=0.65cm]B.south)[below]{$b)$};
   \node(CC) at ([yshift=0.65cm]C.south)[below]{$c)$};
   \node(CD) at ([yshift=0.65cm]D.south)[below]{$d)$};
   \node(CE) at ([yshift=0.20cm]E.south)[below]{$e)$};
   \node(CF) at ([yshift=0.20cm]F.south)[below]{$f)$};
  \end{tikzpicture}
  \caption{$(a,c,e)$ Streamwise and $(b,d,f)$ vertical
           coordinate of the LEV core for the case with $AR=4$ 
           as a function of time during the downstroke. 
           The spanwise sections considered are 
            $y=0.25b$ ($a,b$), 
           $y=0.5 b$ ($c,d$), 
           and $y=0.75b$ ($e,f$).
           Lines as in \ref{fig:LEVskpos_tT250}.
           The colored shaded area indicates the uncertainty in the position of the LEV.
           The black dashed lines in panels $(c)$ and $(e)$ are parallel to $x_c = 0.4u_\infty t$.
  \label{fig:LEVskpos_tT} }
 \end{center}
\end{figure}
%

From the point of view of the characterization of the LEV over flapping wings, 
it is also necessary to address its evolution in time. 
Figure \ref{fig:LEVskpos_tT} shows the time evolution of $x_c$ and $z_c$ for the
case with $AR=4$, at three positions corresponding to 25, 50 and 75\% 
of the span of the wing. 
Near the inboard wing tip (i.e., 25\% of the span, see panels $a$ and $b$), the LEV position 
changes little during the downstroke. The effect of $Q^\prime_{th}$ on the vertical position 
is small, while $x_c$ decreases as the threshold increases, as already discussed in figures 
\ref{fig:LEVskpos_tT250}$b$ and $d$. 
Note that in this spanwise section, the LEV is only detected in the interval 
$0.2 \lesssim t/T \lesssim 0.3$ (i.e., around mid-downstroke) for the highest threshold, 
while it is detected during (almost) the whole downstroke for the lowest threshold. 

More interesting is the evolution of $x_c$ and $z_c$ in the 50 and 75\% spanwise sections. 
During the first half of the downstroke the LEV moves downstream and vertically, with little 
uncertainty and scatter between the different thresholds.
However, both uncertainty and scatter increase considerably around mid downstroke (i.e., $t/T = 0.25$). 
During the second half of the downstroke, $x_c$ moves downstream at a 
roughly constant velocity of about $0.4 u_\infty$ 
(i.e., see black dashed line in figure \ref{fig:LEVskpos_tT}$c$ and $e$).  
Meanwhile, $z_c$ increases and reaches a shallow maximum at a vertical distance from the wing that increases with $y$.
Note that the vertical distance from the LEV core to the wing is relatively small 
($z_c \lesssim 0.5c$ at 75\%, and $z_{c,\max} = 0.63c$ at $76\%$), even if the chordwise motion of the LEV core seems to suggest that its kinematics are somehow detached from the wing's motion.   

\begin{figure}
 \begin{center} 
  \begin{tikzpicture}
   \coordinate(O) at (0.,0.); 
   %
   \node(A) at (O)              {\includegraphics[width=\twhalf,trim=1cm 0cm 2cm 0cm, clip]{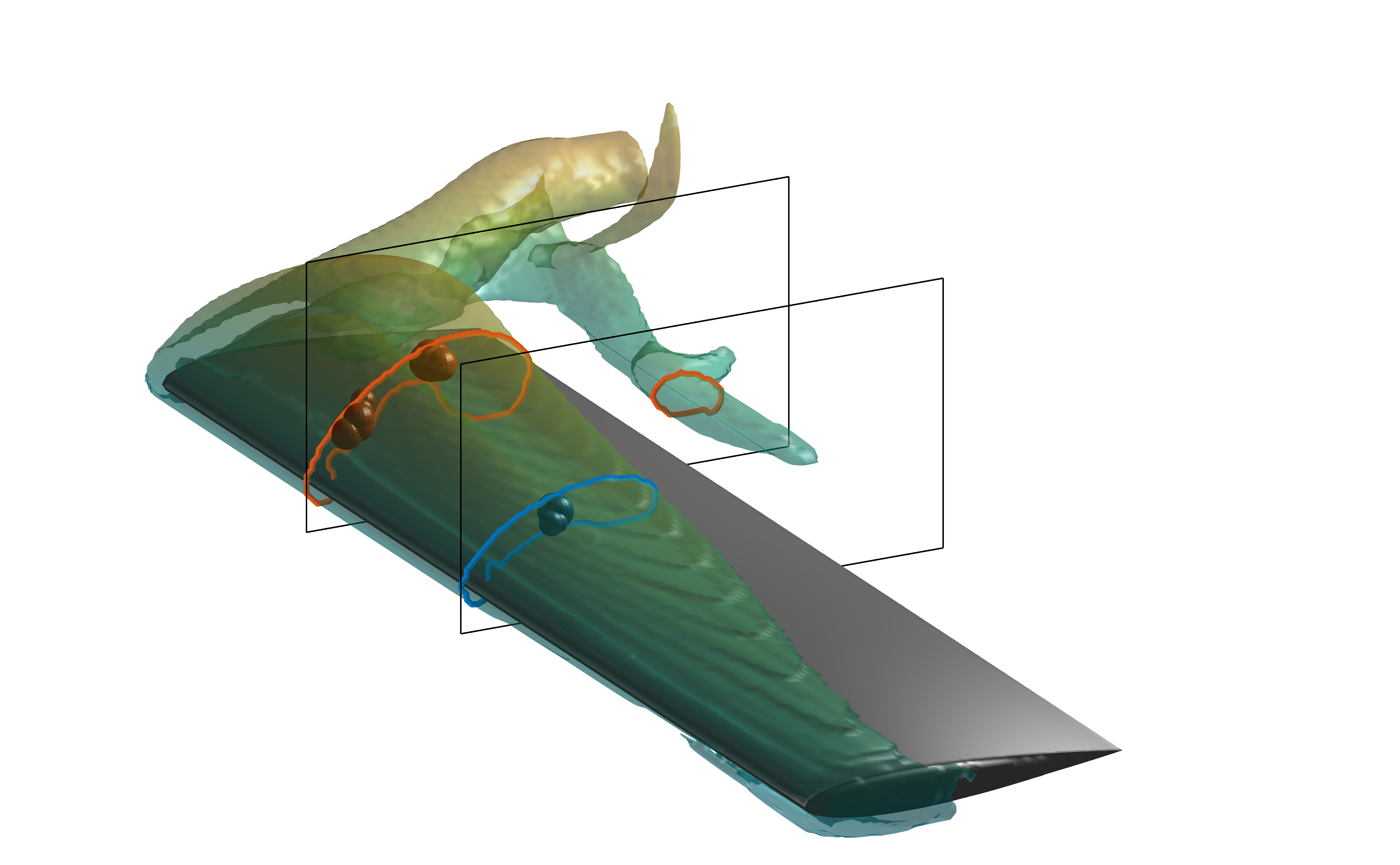}};
   \node(B) at (A.east)[right]  {\includegraphics[width=\twhalf,trim=1cm 0cm 2cm 0cm, clip]{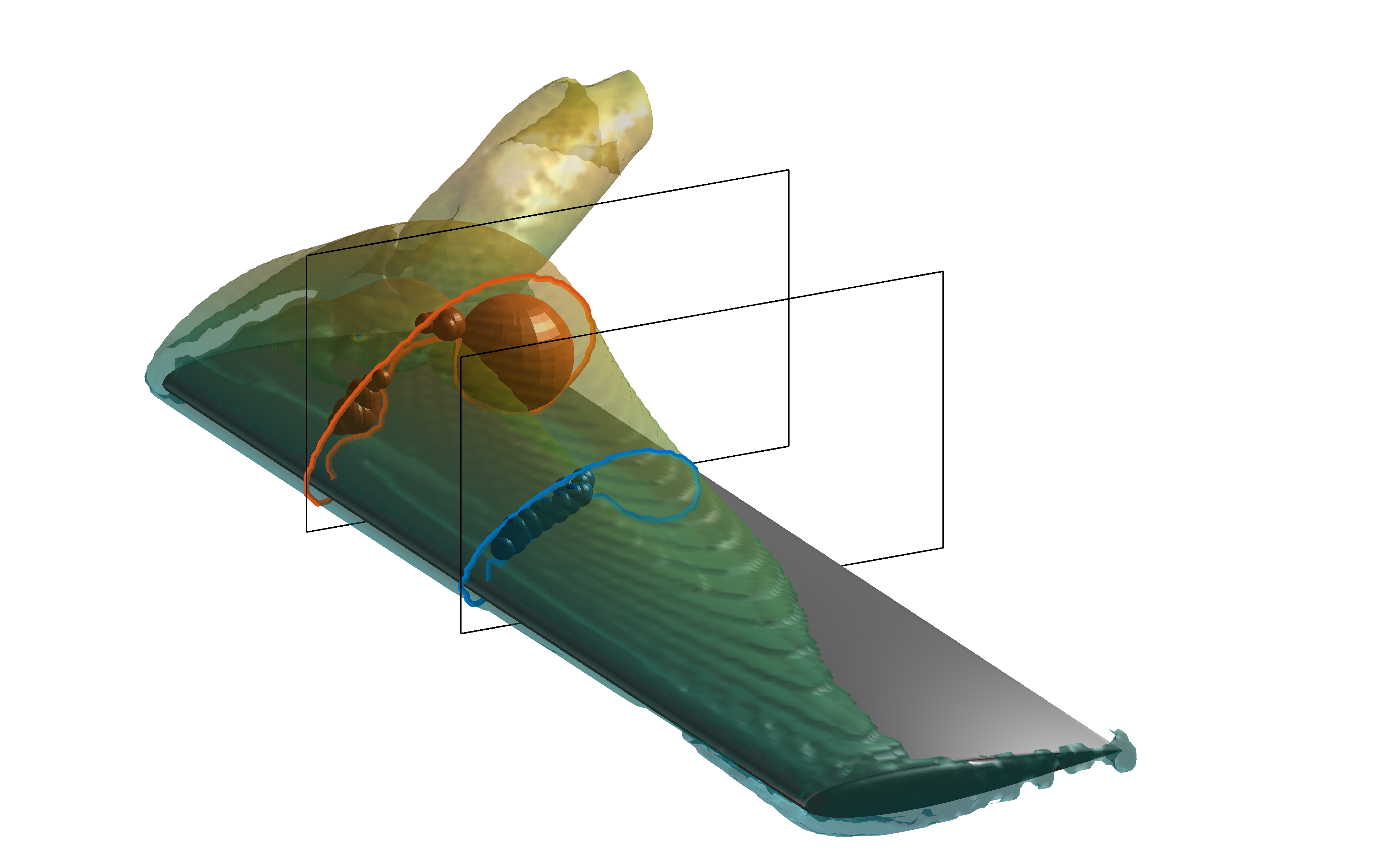}};
   \node(C) at (A.south)[below] {\includegraphics[width=\twhalf,trim=1cm 0cm 2cm 0cm, clip]{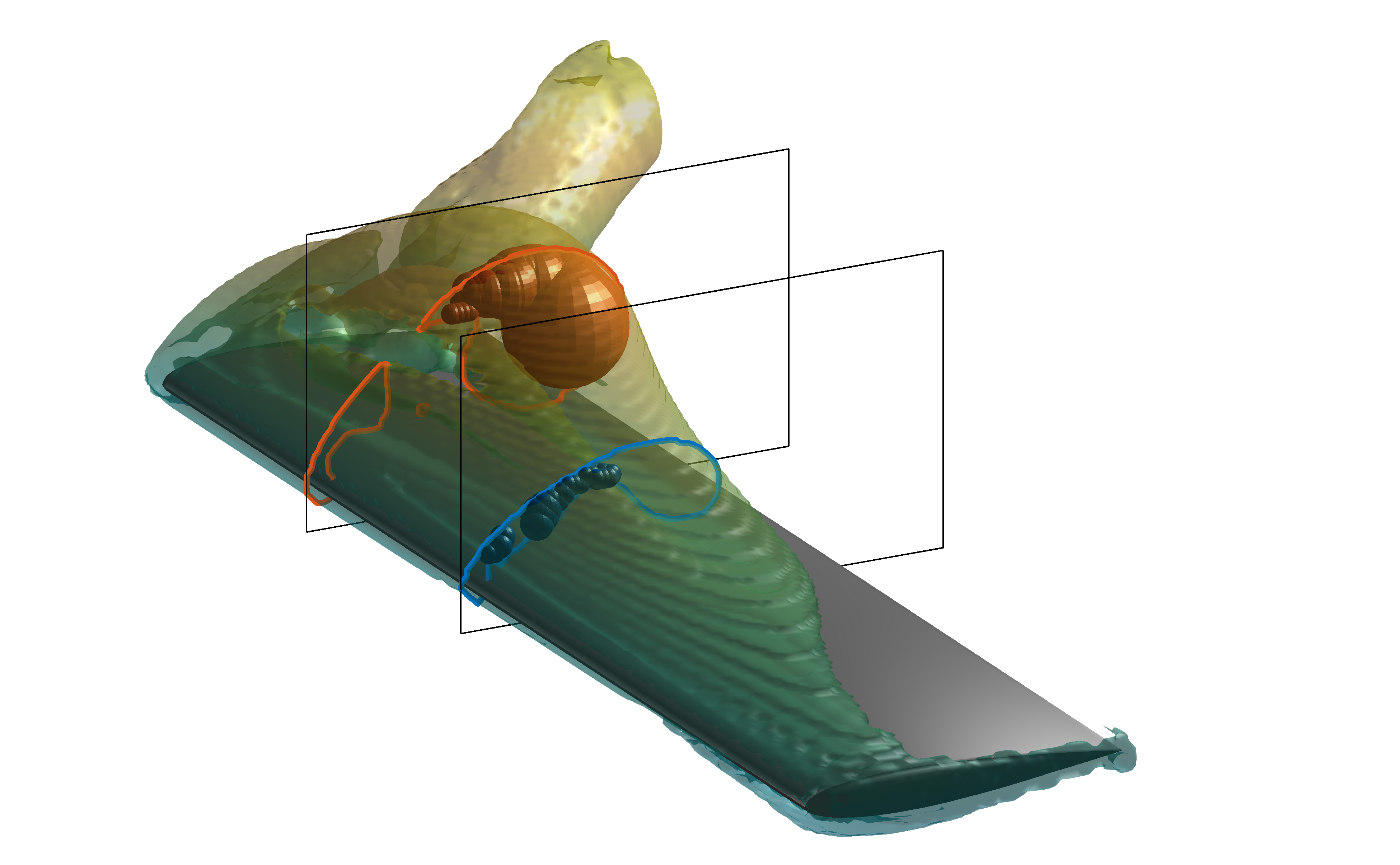}};
   \node(D) at (C.east)[right]  {\includegraphics[width=\twhalf,trim=1cm 0cm 2cm 0cm, clip]{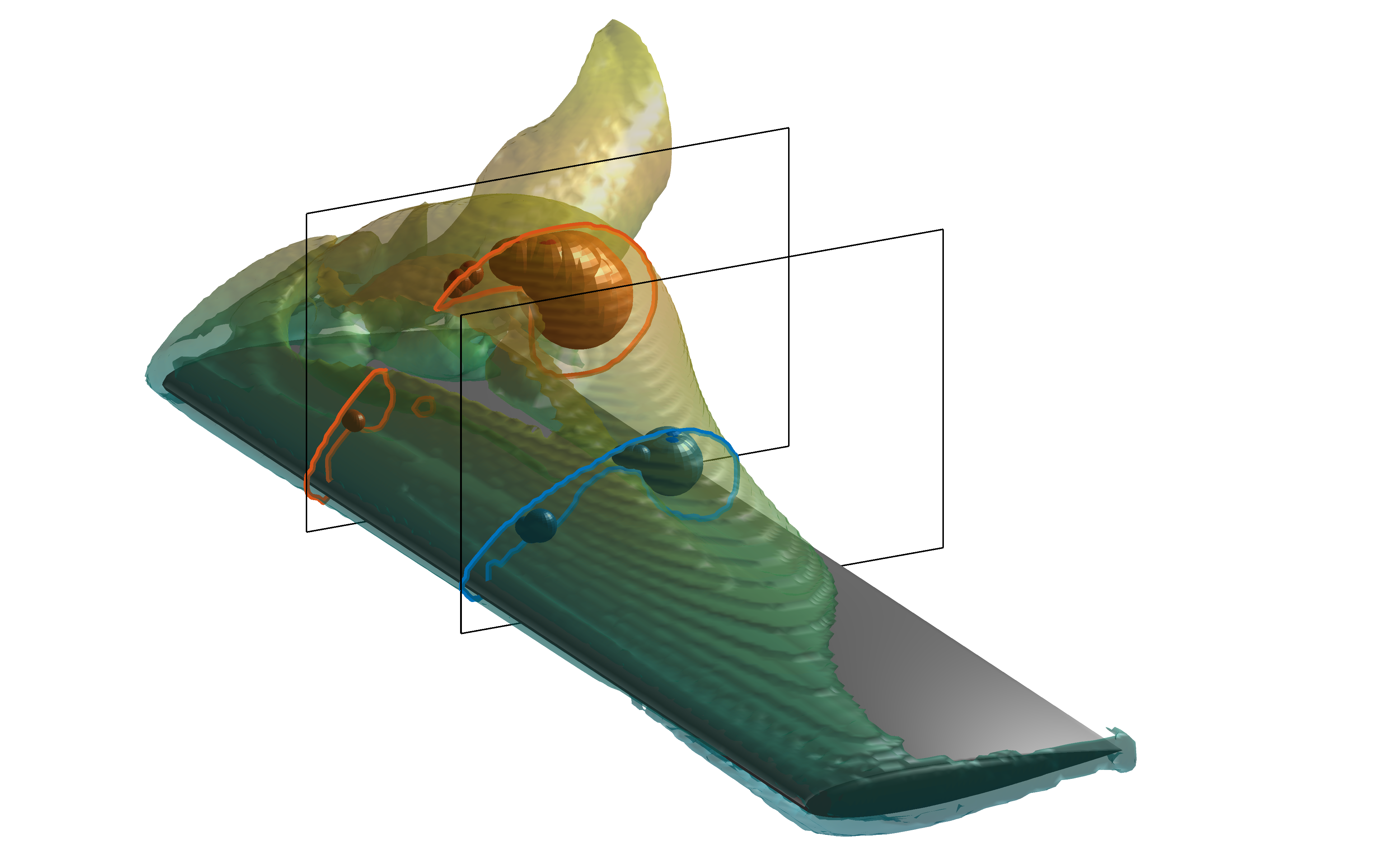}};
   \node(E) at (C.south)[below]  {\includegraphics[width=\twhalf,trim=1cm 0cm 2cm 0cm, clip]{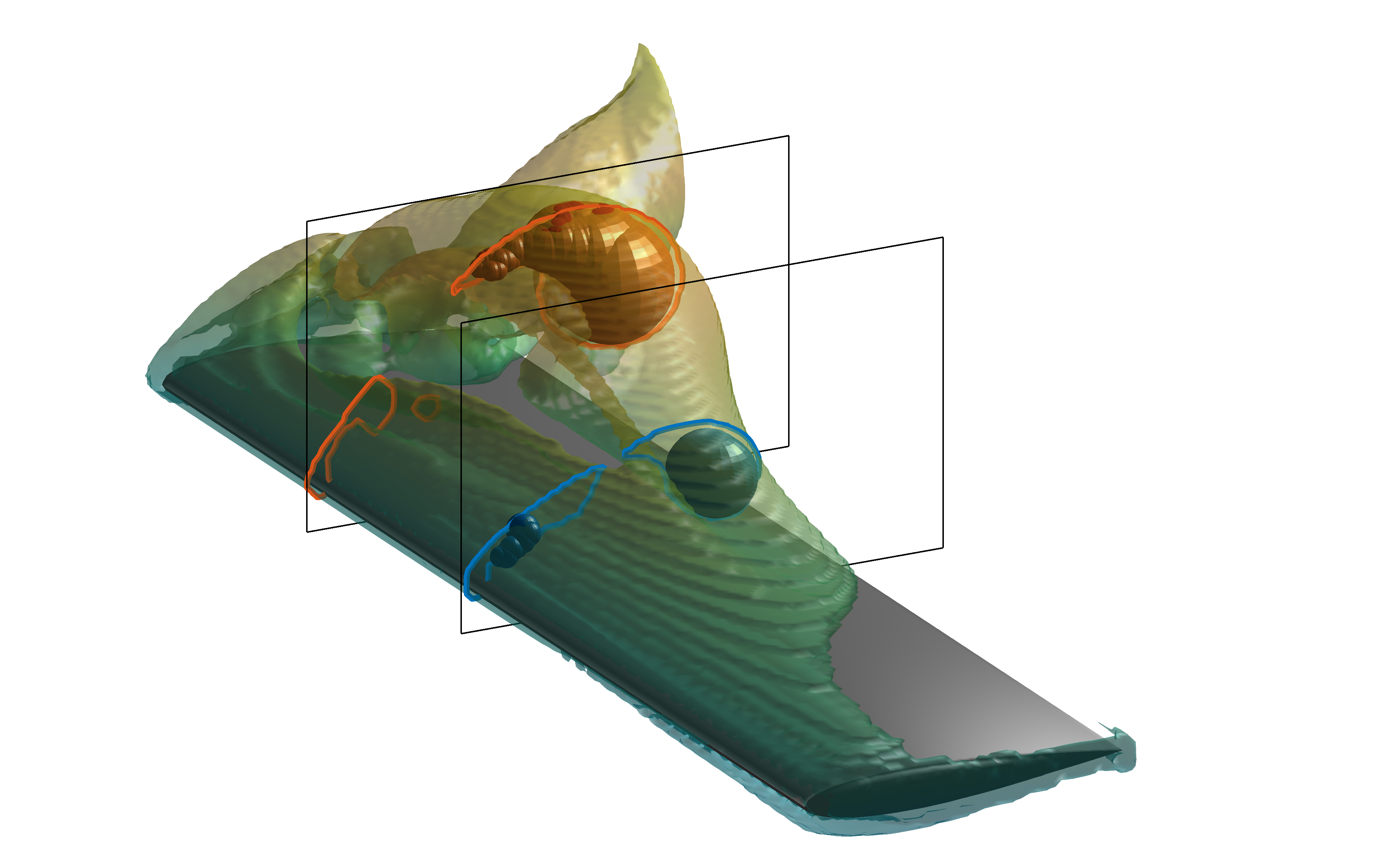}};
   \node(F) at (E.east)[right]  {\includegraphics[width=\twhalf,trim=1cm 0cm 2cm 0cm, clip]{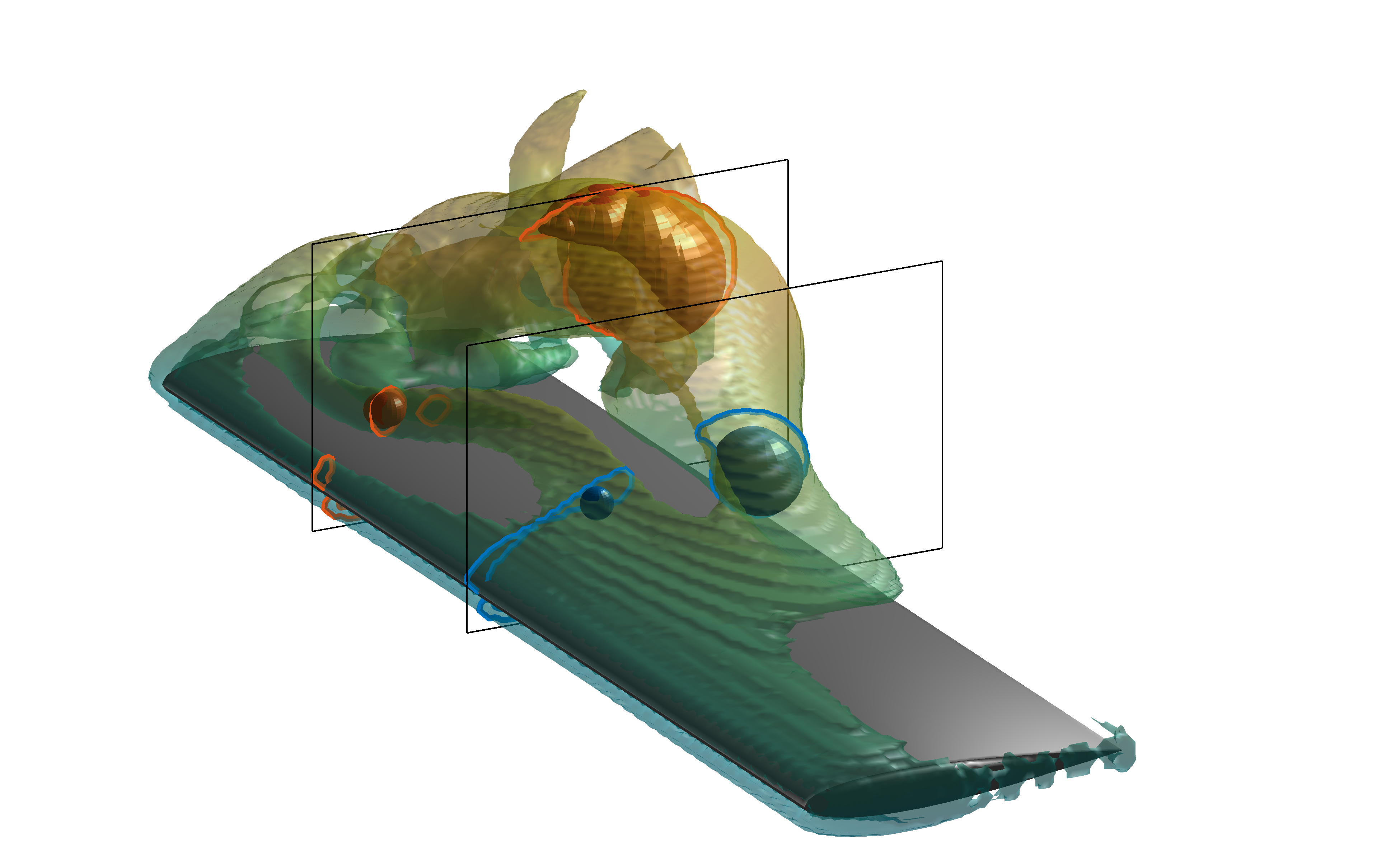}};
   %
   \node(CA) at ([yshift=0.30cm]A.south)[below]{$a)$};
   \node(CB) at ([yshift=0.30cm]B.south)[below]{$b)$};
   \node(CC) at ([yshift=0.30cm]C.south)[below]{$c)$};
   \node(CD) at ([yshift=0.30cm]D.south)[below]{$d)$};
   \node(CE) at ([yshift=0.30cm]E.south)[below]{$e)$};
   \node(CF) at ([yshift=0.30cm]F.south)[below]{$f)$};
  \end{tikzpicture}
  \caption{Isosurfaces of $Q^\prime = 4.0 u_\infty^2/c^2$ for the case of $AR=4$ at
           $(a)$ $t/T = 0.25$, $(b)$ $t/T = 0.31$, $(c)$ $t/T = 0.34$,
           $(d)$ $t/T = 0.38$, $(e)$ $t/T = 0.41$ and $(f)$ $t/T = 0.47$.
           Panels also display the intersection of the planes $y/b = 0.5$ (blue)
           and $y/b = 0.75$ (orange) with the isosurfaces.
           The inscribed spheres associated to the skeleton points at these two 
           spanwise location are also shown. \reva{Isosurfaces of $Q^\prime$ are colored
           with the vertical coordinate of the LEV (i.e., $z/c$) with color transitioning 
           from green to yellow as the distance from the wing's chord line increases.}
  \label{fig:FVA_LEV_separation} }
 \end{center}
\end{figure}
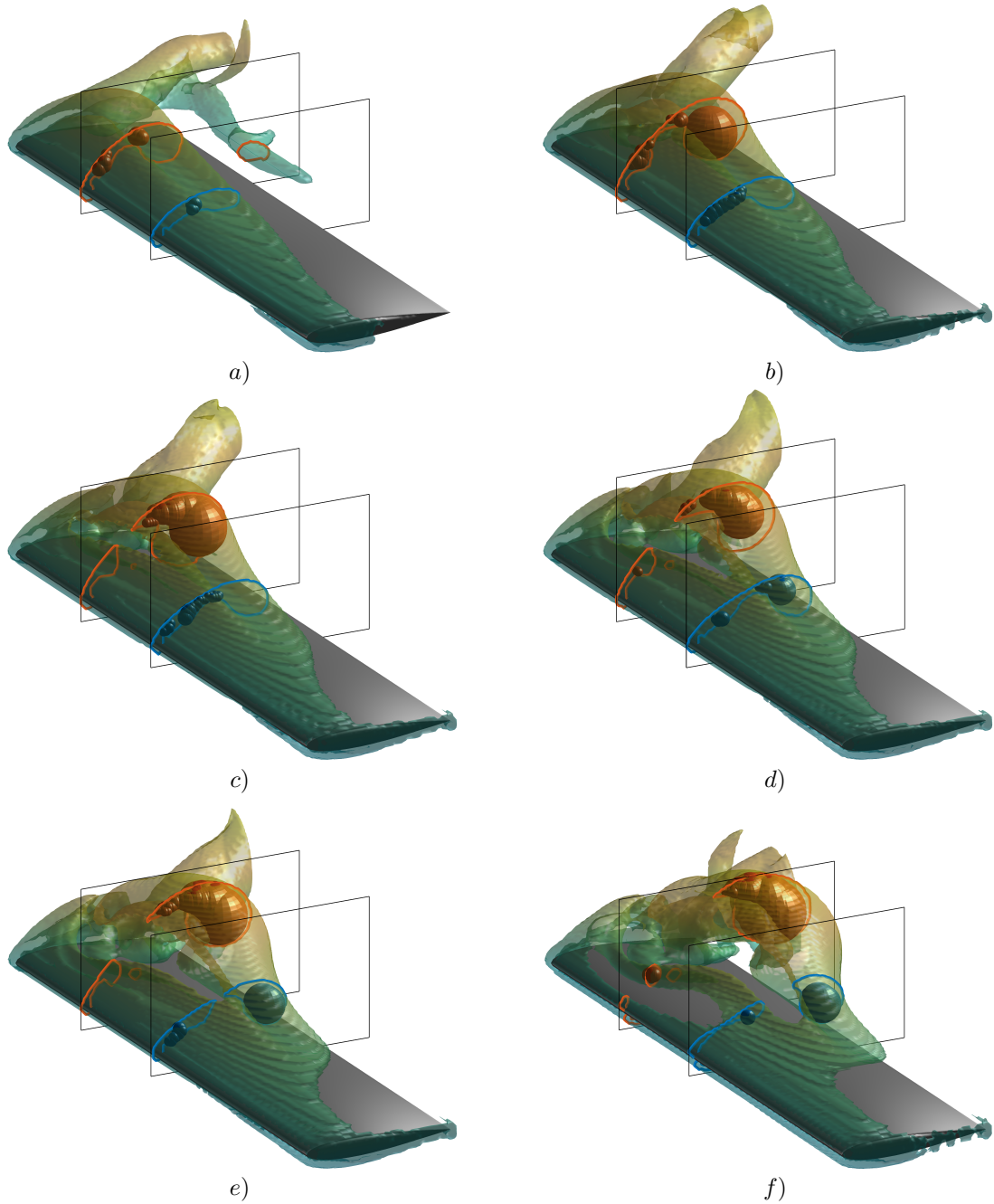

The origin of the uncertainty and the scatter in $x_c$ and $z_c$ during 
the second half of the downstroke is investigated in figure \ref{fig:FVA_LEV_separation}, \reva{and it is related to the skeleton points describing the thin shear layers connecting the LEV core to the leading edge of the wing.} 
This figure shows the points of the skeleton of the LEV in two spanwise bins, corresponding to 
50\% and 75\% of the span of the wing. 
The points are represented by their corresponding inscribed spheres. 
The LEV is represented by the isosurface $Q^\prime = 4u_\infty^2/c^2$ (translucent), as well as the intersection of 
the isosurface with chordwise-vertical planes at the sections 50\% (blue) and 75\% (red) of the wing span.  
At $t/T=0.25$, as already discussed above, the LEV shape is elongated in streamwise direction and 
thin in vertical direction. 
As time increases, the LEV evolves by growing in the downstream part while remaining thin near 
the leading edge. 
Eventually, a bottleneck is produced between the thicker part (downstream) and the thinner part, 
at $t/T\approx 0.3$. 
Somewhat later, pinch off takes place, so that the LEV splits into two 
structures, the first one remaining near the leading edge and the second one traveling downstream.
This phenomenon does not happen simultaneously over the whole span, but rather it starts 
near the outer wing tip and progresses towards the inboard wing tip as time increases.
Thus, pinch off is observed at $t/T \approx 0.34$ at 75\% of the wing span and at 
$t/T \approx 0.4$ at 50\% of the wing span. 
Note that these times are dependent on the particular $Q^\prime_{th}$ selected for the visualization. 

In summary, the LEV evolution can be described as an elongated structure in the spanwise direction
that grows and splits with the shape of the letter ``y'', 
similar to that observed in previous works \citep{harbig2013,jardin2014}. 
The weaker (i.e., smaller) leg remains close to the leading edge, and eventually disappears at the end of the stroke. 
The strongest (i.e., larger) leg of the vortical structure remains relatively close to the wing surface, traveling downstream at a roughly constant velocity.
%
Note that near the end of the downstroke, the LEV branch that remains closer to the leading edge splits again (see figure \ref{fig:FVA_LEV_separation}$f$).

Figure \ref{fig:FVA_LEV_separation} also shows that the uncertainty in $x_c$ and $z_c$ observed in figure \ref{fig:LEVskpos_tT} 
for the times and spanwise sections where the LEV is split 
is associated to the presence of points of the skeleton of the LEV in both branches 
of the y-shaped LEV. 
On the other hand, the effect of $Q^\prime_{th}$ on the time of the pinch off results in the aforementioned scatter in the lines in figures \ref{fig:LEVskpos_tT}$c$ to $f$. 

Although not shown here, a similar picture is obtained for the $AR=2$ wing: the development of a y-structure in the LEV, with the downstream branch of the vortex being advected downstream at a roughly constant velocity (i.e., $0.4 u_\infty$) while its vertical coordinate relative to the wing remains within $z_c \lesssim 0.5 c$).

\def\tmpa{FVA_LEV_separation_relQ_noxth}
\def\tmpb{2W-AR4-R000_Qth4000}

One of the most elusive features of the dynamics of the LEV is the precise 
definition of its separation (and/or breakdown), 
and the effect that such separation might have in the 
aerodynamic forces over the wing \citep{lentink2009b,jardin2014,birch2004,ozen2012}.
The results obtained from force decomposition algorithms in 2D configurations 
\citep{chang1992, martin2015vortex,moriche2017b,menon2021quantitative}
suggest that the effect of the vortices on the lift are important provided that the vortices are sufficiently close to the wing, roughly within one chord from the wing. 
From that point of view, the effect of the LEV on the forces of the present configurations 
should still be relevant, even while the LEV core is being advected downstream at a roughly 
constant velocity during the second half of the stroke. 
Hence, the methodology proposed here to quantify the LEV core position and physical properties 
is used next to evaluate the evolution of the circulation of the LEV vortex and its effect on the 
aerodynamic forces on the wing. 

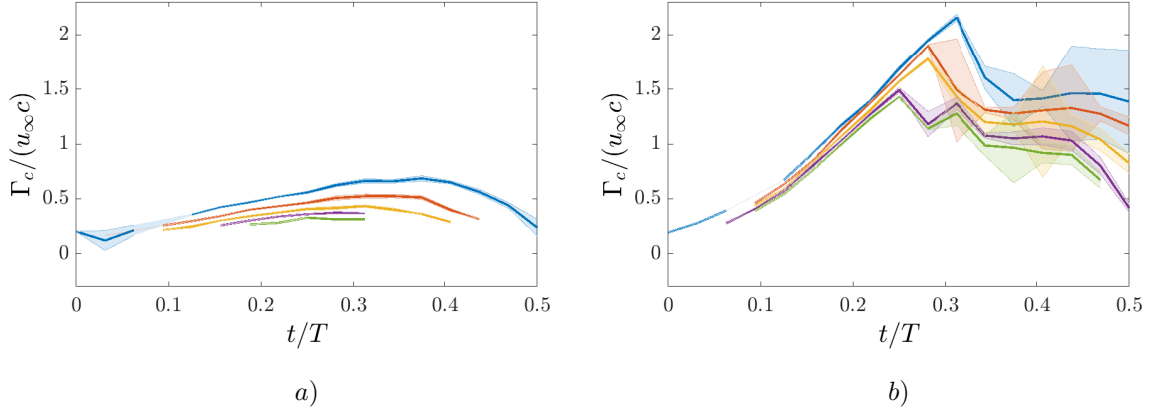
\begin{figure}
 \begin{center} 
  \begin{tikzpicture}
   \coordinate(O) at (0.,0.); 
   %
   \node(A) at (O) {\includegraphics[width=\twhalf,trim=0cm 0cm 0cm 0cm, clip]{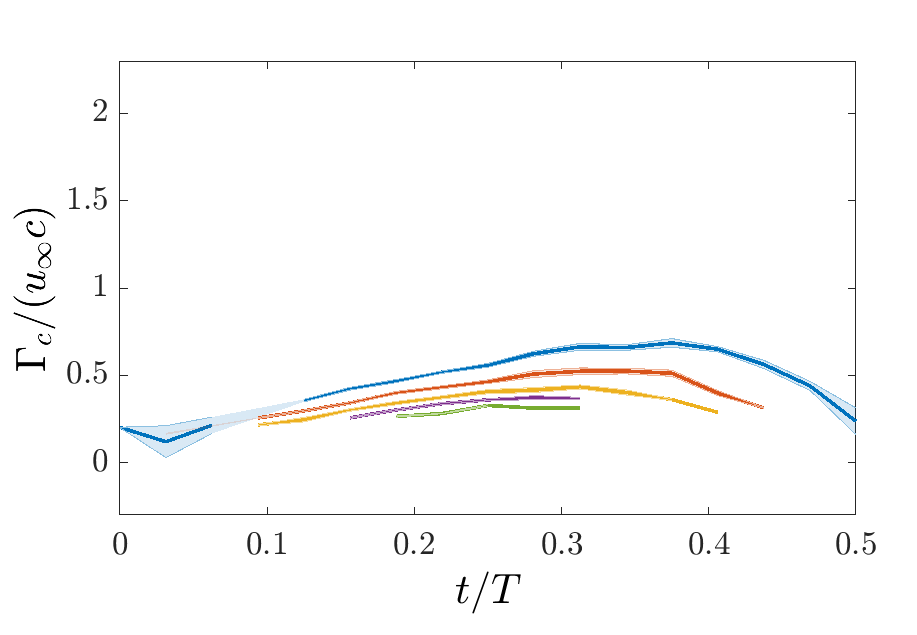}};
   \node(B) at (A.east)[right] {\includegraphics[width=\twhalf,trim=0cm 0cm 0cm 0cm, clip]{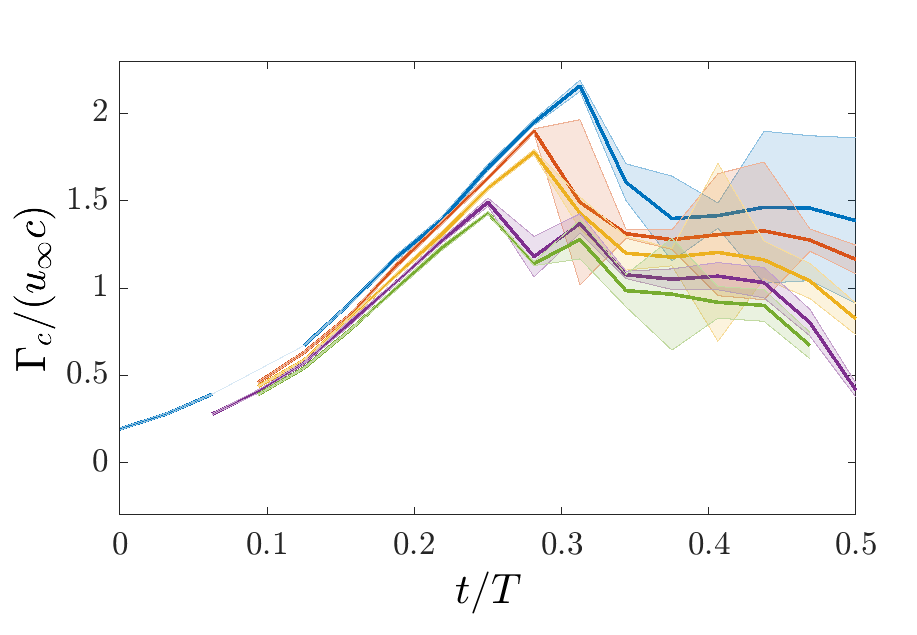}};
   %
   \node(CA) at ([xshift=0.25cm,yshift=0.00cm]A.south)[below]{$a)$};
   \node(CB) at ([xshift=0.25cm,yshift=0.00cm]B.south)[below]{$b)$};
  \end{tikzpicture}
  \caption{Circulation on the LEV core ($\Gamma_c$) as a function 
  of time during the
           downstroke at $(a)$ $y = 0.25b$ and $(b)$ $y = 0.75b$.
           Lines as in figure \ref{fig:LEVskpos_tT250}.
           The colored shaded area indicates the corresponding uncertainty.
  \label{fig:Gamma_tT} }
 \end{center}
\end{figure}

Figure \ref{fig:Gamma_tT} shows the local circulation of the LEV core for case $AR=4$, at the spanwise positions 25\% and 75\%. 
Near the inboard wing tip, figure \ref{fig:Gamma_tT}$a$, the circulation increases smoothly during most of the downstroke, peaking at times well past the mid-downstroke (i.e., when the vertical speed of the wing and the effective angle of attack is maximum). As expected, the value of the threshold limits the time interval when the LEV core is detected, as well the value of the local circulation. 
On the other hand, as shown in figure \ref{fig:Gamma_tT}$b$, 
the local circulation at the 75\% spanwise section increases steadily during the downstroke, to suddenly reach a more or less constant value after a slight overshoot. 
Comparison of figures \ref{fig:Gamma_tT}$b$ and \ref{fig:LEVskpos_tT}$e$ shows that 
the time at which $\Gamma_c$ reaches a plateau roughly coincides with 
the advection of the LEV core at a constant velocity
(i.e., when $x_c$ grows linearly with time in figure \ref{fig:LEVskpos_tT}$e$). 
Not surprisingly, the time when $\Gamma_c$ reaches a plateau 
and the magnitude of the overshoot depend on $Q^\prime_{th}$. 
Also, the uncertainty in $\Gamma_c$ during the overshoot and subsequent plateau increases, 
probably due to the y-shape of the LEV and the difference in $\Gamma_s^k$ for points in the 
upstream or downstream branches of the y-shaped LEV (see figure \ref{fig:FVA_LEV_separation}). 
Although not shown, the evolution of $\Gamma_c$ at the 50\% spanwise position is qualitatively similar to that obtained at 75\%.

Conceptually, 
figures \ref{fig:LEVskpos_tT} and \ref{fig:Gamma_tT} suggest that the evolution of the LEV has two distinct phases. During (roughly) the first half of the downstroke, the LEV develops and grows increasing its circulation. 
Then, the LEV splits, and its downstream section is advected towards the wake while  keeping its circulation approximately constant. 
The picture is very similar to that reported by \cite{jardin2014} in revolving 
wings using 2D visualizations, with values for the peak and plateau of the circulation 
of the same order of magnitude as those reported here. 

\def\tmpa{LEV_stats_yw_tT_th_influence_relQ_noxth}
\def\tmpb{2W-AR4-R000}

\begin{figure}
 \begin{center}
  \begin{tikzpicture}
   \coordinate(O) at (0.,0.);
   %
   \node(A) at (O) {\includegraphics[width=\twhalf]{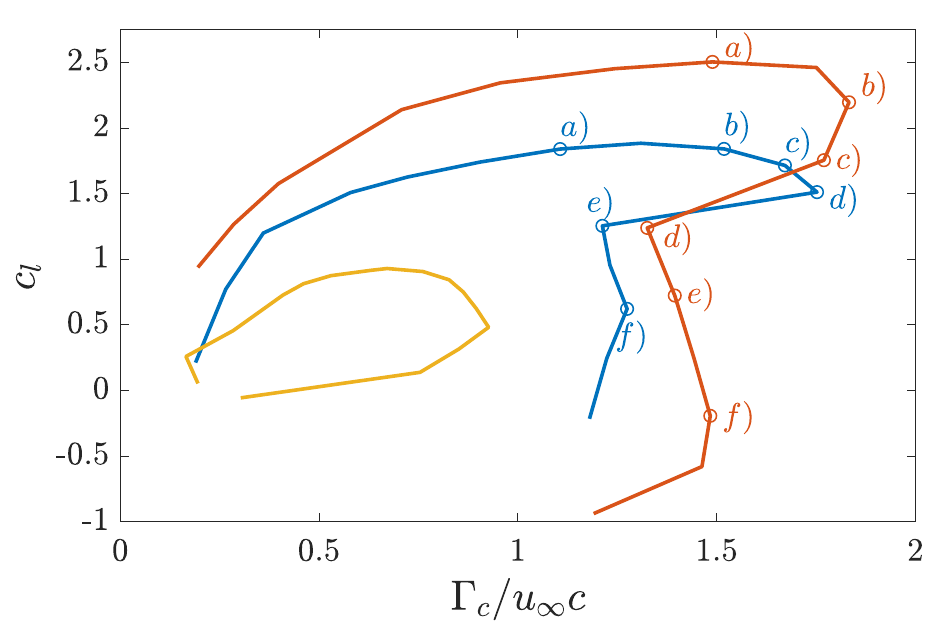}};
   \node(B) at (A.east)[right] {\includegraphics[width=\twhalf,trim=0cm 0cm 0cm 0cm, clip]{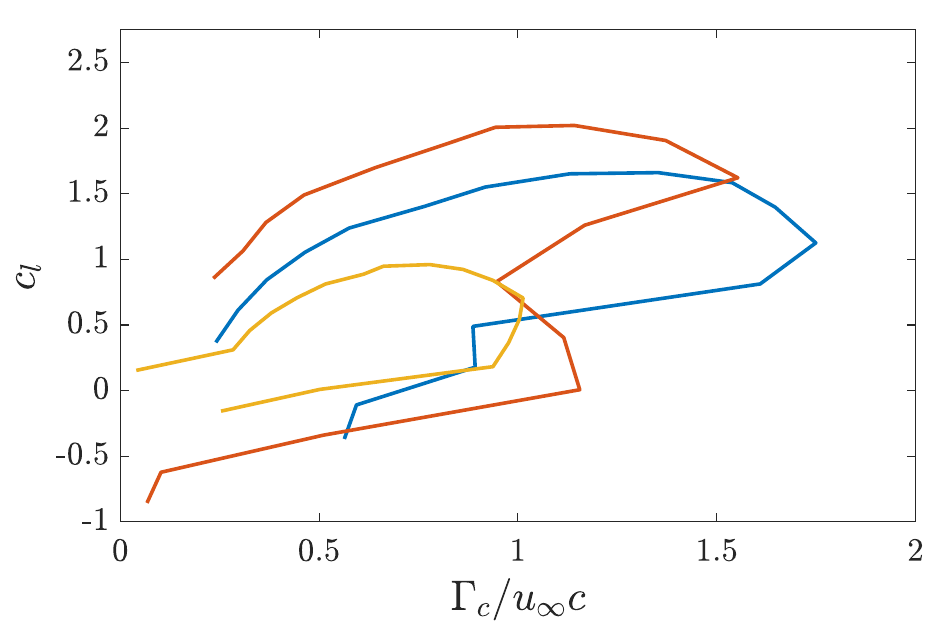}};
   %
   \node(CA) at ([xshift=0.50cm,yshift=0.00cm]A.south)[below]{$a)$};
   \node(CB) at ([xshift=0.50cm,yshift=0.00cm]B.south)[below]{$b)$};
  \end{tikzpicture}
  \caption{Evolution of the sectional lift coefficient, $c_l$,
           vs. the circulation of the LEV core, $\Gamma_c$, during the downstroke. Lines correspond to spanwise sections 
           $y = 0.25b$ (yellow), $y = 0.5b$ (blue) and $y = 0.75b$ (orange).
           ($a$) case with $AR=4$, 
           ($b$) case with $AR=2$. 
           The letters corresponds to the panels (i.e., times) shown in figure   \ref{fig:FVA_LEV_separation}.
  \label{fig:cnzvsGamma} }
 \end{center}
\end{figure}
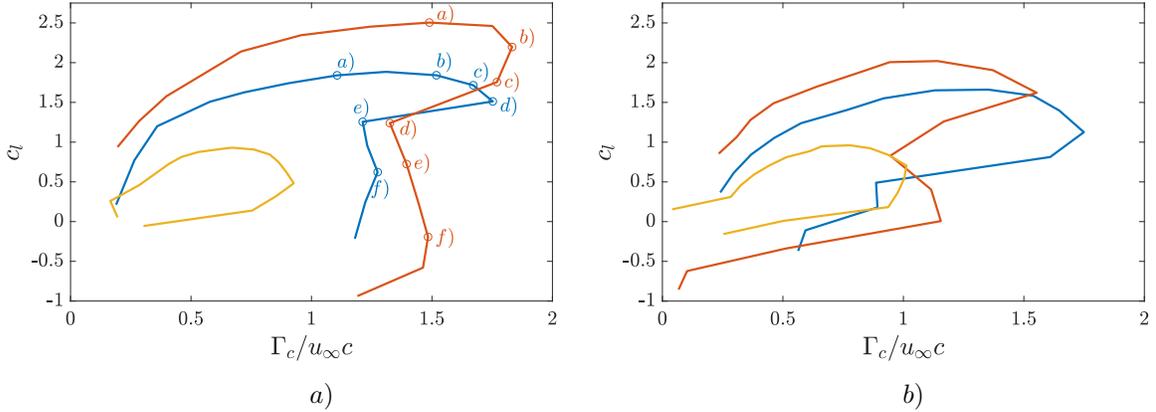

Finally, figure \ref{fig:cnzvsGamma} evaluates the link between the 
local circulation of the LEV core and the local aerodynamic force, 
characterized here with the sectional lift coefficient 
\begin{equation}
c_{l}(y) = \frac{l(y)}{1/2 \rho u_\infty^2 c}, 
\end{equation}
where $l(y)$ is the sectional lift per unit span, i.e. the resultant of the aerodynamic forces in the vertical direction (inertial system of reference) at a spanwise section ($y$) . 
The figure shows $c_l$ as a function of $\Gamma_c$ during the downstroke 
for cases with $AR=4$ (figure \ref{fig:cnzvsGamma}$a$) and 
with $AR=2$ (figure \ref{fig:cnzvsGamma}$b$). 
Three spanwise sections are plotted with different line colors, 25\% (blue), 50\% (yellow) and 75\% (red). 
The local circulation is computed for $Q^\prime_{th} = 4u_\infty^2 /c^2$, although similar plots are obtained for other thresholds. 
For the case with $AR=4$ (figure \ref{fig:Gamma_tT}$a$), the labels on the 
lines for 50\% and 75\% of the span corresponds to the labels of figure 
\ref{fig:FVA_LEV_separation}, so that time increases in clockwise direction
for all loops in the figure. 

\def\tmpa{cz_n_tot_Gamma_yw5075_relQ_noxth}
\def\tmpb{2W-AR4-R000_Qth4000}
\def\tmpc{2W-AR2-R000_Qth4000}

Focusing first in the case with $AR=4$,  figure \ref{fig:cnzvsGamma}$a$
shows that the maximum $c_l$ (which occurs shortly after mid-downstroke, $t/T \gtrsim 0.25$) is obtained before the peak value of $\Gamma_c$.
Indeed, between the maximum $c_l$ and the maximum $\Gamma_c$, the local circulation still increases by about 20-30\%. 
At the 25\% spanwise section, the evolution of both $c_l$ and $\Gamma_c$ is smooth.
However, and consistently with the time histories shown in figure \ref{fig:Gamma_tT}, 
at the spanwise sections 50\% and 75\% 
there is a sudden decrease in $\Gamma_c$ just after its maximum. 
During the subsequent plateau in $\Gamma_c$, the value of the sectional lift coefficient decreases monotonically, as the LEV core is advected downstream.  

Similar observations can be made for the case with the smaller aspect ratio, shown in figure \ref{fig:cnzvsGamma}$b$. 
In this case, the reduced $AR$ results in a less clear plateau of $\Gamma_c$, although the main characteristics observed in figure \ref{fig:cnzvsGamma}$a$ can still be identified: maximum $c_l$ while $\Gamma_c$ is still growing, sudden decrease of $\Gamma_c$ after its maximum for 50\% and 75\% spanwise sections, etc.

%
\section{Conclusions}
\label{sec:conclusions}

The LEV is one of the most important unconventional aerodynamic mechanisms
providing lift augmentation in flapping wing aerodynamics.
The LEV is essentially an elongated structure that grows and evolves, and during
the course of the flapping oscillation can present complicated shapes, including changes in the topology (i.e., splitting).
In this work, we have proposed a methodology to analyze the LEV that 
takes into account these complexities
aiming to provide a quantitative description of the LEV.
The first step involves the identification of the vortical structures surrounding the wing. 
The identification of the structures has been done 
employing an isosurface of the second invariant of the
velocity gradient tensor, $Q$, but, in principle,
it can be done with any of the methods existing in the literature \citep{chakraborty2005}.
The second step consists of the identification of the skeleton or core of the LEV using a thinning algorithm also available in the literature \citep{lee1994,kerschnitzki2013}, 
\reva{and the definition of the local vorticity vector in each point of the skeleton.
The third step consist on the discrimination of the LEV 
from the remaining vortical structures, which is done using geometrical considerations.} 
The \reva{fourth} step consists of the computation of relevant flow quantities along the LEV core.
This is done by averaging the flow quantities within 
planes perpendicular to the local vorticity vector, which is used here to define the local direction of the LEV core. 
We have presented the results as a function of the wing span, and for this purpose, the
results have been additionally averaged using bins along the span. 
This last averaging procedure is not strictly necessary and other alternatives for the presentation of the results are possible, in general.

\reva{It should be noted that our methodology (specifically, steps 1, 2 and 4) can be used to characterize any vortical structure, including particularly complex three-dimensional ones, even if in this paper we have restricted ourselves to the LEV.  
Traditional methods rely on user-defined slicing to calculate metrics (e.g., vorticity, circulation) integrated in vortex regions, assuming known vortex orientation (e.g., LEV direction parallel to the wingspan), which is not always the case. 
In contrast, our algorithm automatically detects optimal slicing aligned with the vortical structure, enabling a more accurate integration of quantities of interest along the vortex core. }

As an illustration of the methodology, we have analyzed flow data corresponding to 
a pair of wings performing a flapping motion as they fly forwards at constant speed. 
Two aspect ratios have been considered, namely $AR = 2$ and $AR=4$.
For this particular configuration, we have provided a geometrical characterization of the 
LEV, by tracking, as a function of time, 
the vortex core along the wing span during the downstroke. 
We have shown that near the inboard wing tip the position of the LEV core changes little 
during the downstroke.
However, the part of the LEV located beyond the half-span evolves significantly during the downstroke.
During the first half of the downstroke, the LEV core moves downstream and vertically, at
a slow but increasing pace. 
During the second half of the downstroke the convection velocity in streamwise direction 
is rather constant suggesting that detachment from the wing has begun. 
However, 
the relative distance to the wing is still relatively small, of the order of half a chord.
Also during this second half of the downstroke, we have shown that the uncertainty in the position of the LEV core is large. 
Flow visualizations have shown that the large uncertainty is associated to the subsequent splitting of the LEV, first developing a y-shape and ending with three individual cores at the end of the downstroke. 

In addition to the geometrical characterization of the LEV, 
we have analyzed the pressure inside the LEV,
the velocity and vorticity components along the LEV core, and with particular emphasis, 
the local circulation of the LEV. 
Near the inboard wing tip, the local circulation increases smoothly
during most of the downstroke.
Near the outboard wing tip the local circulation grows linearly during the
first half of the downstroke reaching a plateau during the second half of the downstroke.
Thus, the LEV evolution can be conceptually divided in two phases. 
First, the LEV develops and grows increasing its circulation smoothly. 
Approximately at mid-downstroke the LEV starts splitting and its downstream part
is advected towards the wake while keeping its circulation rather constant.
To close the article, we have explored the link between the sectional lift on the 
wing and the local circulation. 
This information, when obtained for a sufficiently large database, can lead to improvements
of simplified models of the aerodynamic force for flapping wing configurations.
Such improvements could have a significant impact towards the systematic design of MAVs.

Finally, note that we have proven the robustness of the methodology employed by analyzing the variation of the results with the threshold selected 
for the identification of the vortical structures.
The impact of the threshold has shown to be minor not influencing the observed trends.
We believe that the present work is a step towards a more complete characterization
of the leading edge vortex. 
This is a difficult task because of the complexity of the LEV structure and its time evolution. 
The methodology presented here consists of several steps that can be improved and we hope that the present work stimulates the discussion on this topic.

\section*{Availability of data and materials}
The datasets analysed during the current study are available from the corresponding author on reasonable request.

\section*{Competing interests}
The authors declare that they have no competing interests

\section*{Funding}
This work was supported by grant DPI2016-76151-C2-2-R (AEI/FEDER, UE).

\section*{Authors' contributions}
AG: Acquisition, analysis and interpretation of data. Software. Writing: original draft.
MGV: Conception and design of the work. Writing: revision.
OF: Conception and design of the work. Writing: revision.

\section*{Acknowledgements}
Nothing to declare.

\bibliographystyle{jfm}
\bibliography{biblio}

\end{document}